\documentclass[12pt]{article}
\usepackage{epsfig}
\def\m{m_\pi}

\def\bfm#1{{\mbox{\boldmath $#1$}}}
\def\Tr{{\rm Tr}}
\def\intt#1{\int \frac{d^3#1}{(2\pi)^3}}
\begin{document}
 \author{ Rinaldo Cenni $^{1}$, Jishnu Dey $^{2}$ and Mira Dey $^{3}$}
 \title{Nucleons and Isobars at finite density ($\rho$) and temperature 
($T$)}
\maketitle
\noindent
{\it  $^{1)}$I.N.F.N., 
Sezione di Genova and Dipartimento di Fisica, Universit\`a di Genova, via 
Dodecaneso, 33-16146 Genova,  Italy} \\  
{\it $^{2)}$Abdus Salam ICTP, Trieste, Italy and Azad Physics Centre, \\ 
Dept. of Physics, 
Maulana Azad College, Calcutta 700 013, India}\\
{\it $^{3)}$Department of Physics, Presidency College, 
Calcutta 700 073,India; an associate member of 
Abdus Salam International Centre of Theoretical physics, Trieste, Italy}
\vskip 0.2cm\noindent
{  $^{\dagger}$ All communications to 
cenni@ge.infn.it and deyjm@giascl01.vsnl.net.it
\vskip 2 cm
\begin{abstract}
The importance of studying matter at high $\rho$ increases as more 
astrophysical data 
becomes available from  recently launched spacecrafts. The importance of 
high $T$ studies derives from heavy 
ion data. In this paper we set up a formalism to study the
nucleons and isobars with long and short range potentials
non-pertubatively, bosonizing and expanding semi-classically the Feyman
integrals up to one loop. We address the low density, finite T problem 
first, the case relevant to heavy ion collisions, hoping to adresss the
high density case later. Interactions change 
the nucleon and isobar numbers at different $\rho$ and $T$ non-trivially.  
\end{abstract}
\newpage
\section{Introduction}

Isobars ($\Delta $'s) play a very important role in nuclear physics. Two pion
exchange with a $\Delta $  intermediate state is known to produce
the intermediate range attraction between nucleons\cite{MaHoEl-87}. 
Many attempts have been
made to extract the effect of  $\Delta$  intermediate states in nuclear
matter dynamics, and microscopical calculations have shown
the overwhelming relevance they have in binding of nuclear matter 
 \cite{MaHoEl-87,CeCoDi-85}. 

$\Delta $ is the next excited state in the nucleon spectroscopy with a large
degeneracy factor ( $4 \times 4$ for spin and isospin). 
Paradoxically however not too much efforts have been devoted to extract
the amount of $\Delta$ population in the nuclear matter ground state
\cite{An-al-79,CeCoLo-89}. It is on the other hand conceivable that the
small numbers found so far (Ref.  \cite{An-al-79} clearly overestimates
the $\Delta$ component of the nuclear ground state) suggest that the presence 
of $\Delta$'s could at most induce rather small effects in the ground state
properties. Only recently use of modern $N$-$N$ interaction has renewed
the interest in this field \cite{FrKaMuPo-01}, but still dynamical aspects 
like spectral functions and response functions deserve further investigations
even taking advantage of the modern computing facilities.

Even disregarding, however, the relevance of traditional nuclear matter 
calculations, new achievements both on the astrophysical sector and 
on the side of ultra-relativistic heavy-ions collisions suggest to explore
the properties of the hadron matter at finite (high) temperature.
One should keep in mind in fact that in the latter case a low density hadronic
matter is expected to be the relic of a phase transition to and from a
quark-gluon plasma phase. The knowledge of the nuclear matter properties
at low density and high temperature is thus linked (maybe not in a simple way)
to the signals that such a  phase transition has indeed occurred.
But such conditions seem to be much more sensitive to the presence of
$\Delta$'s,  as real $\Delta$'s can be produced.

This has already been established by calculations where 
 non-interacting nucleon and 
 isobar excited states are embedded in a thermal bath ( see for example
Bebie et al. \cite{BeGeGoLu-92} and Dey et al. \cite{DeKrToFr-94}).
Interactions between nucleons ($N$ in short) and $\Delta$'s being strong,
it is very difficult to handle. A non-perturbative treatment is mandatory.

The framework we use in this paper
to deal with a system of interacting nucleons and isobars 
is provided by the  path integral technique developed by one of us and his
co-workers \cite{AlCeMoSa-87,AlCeMoSa-88, CeSa-94, CeCoSa-97}. 
It involves Bosonisation arising from
integration over the fermionic fields.

For non-interacting fermions the methodology is simple. However, for long
range one pion exchange potential ( OPEP ) or short range correlations
(SRC in short) non-perturbative semi-classical approximation schemes are
necessary. This provides a mean field. Higher order corrections can be put
onto it.

The method is described in short in the next section. The nucleon and the
isobar, being heavy particles can be treated non-relativistically.
This allows the Fermi integrations and the Pauli principle to be dealt with.
As the density , $\rho$, increases the baryons are packed more closely so
that the Pauli principle is expected to play a crucial role.

\section{The partition function for the Nucleon interacting with the Pion}

To exemplify, we consider a system of nucleons interacting with pions. OPEP
potential between nucleons and pions is written as
\begin{equation}
V_\pi({\bf q})=\Gamma_i^{(1)}{\cal V}_\pi({\bf q})\Gamma_i^{(2)}
\label{nd1.1}
\end{equation}
where
\begin{equation}
{\cal V}_\pi({\bf q})=-\frac{f^2_{\pi NN}}{\m^2}
\frac{{\bf q}^2}{{\bf q}^2+\m^2}
\label{nd1.2}
\end{equation}
and
\begin{equation}
\Gamma_i^{(1)}=\bfm \sigma_1\cdot \hat{\bf q} \tau_i\;.
\label{nd1.3}
\end{equation}
The partition function in the grand canonical ensemble, in terms of path
integrals, reads
\begin{equation}
Z=e^{-\beta \Omega}=\Tr e^{-\beta(H-\mu N)}=\int
D[\psi^\dagger(x),\psi(x)]e^{-\int_0^\beta d\tau \left[H(\tau)-
\mu N(\tau)\right]}
\label{nd5}
\end{equation}
where $H\,, \mu \, $ and $N$ are the Hamiltonian , the chemical potential and
the number operator respectively.

For non-interacting nucleons we can  easily evaluate the partition
function in terms of the Fourier transforms for the fields  
$\psi({\bf q},\omega_n)$ :
\begin{equation}
Z_0=Z=\int D[\psi^\dagger,\psi] e^{\sum_n \int d^3q
\psi^\dagger(-{\bf q},-\omega_n){\cal G}^{-1}_0(q,\omega_n)\psi({\bf 
q},\omega_n)}
\label{nd6}
\end{equation}
where ${\cal G}_0(q,\omega_n)$  is the free Green's function at finite
temperature :
\begin{equation}
{\cal G}_0(q,\omega_n)=-\frac{\strut 1}{\strut\frac{\strut q^2}{\strut
2M}-\mu-i\omega_n}
\label{nd7}
\end{equation}
with
\begin{equation}
\omega_n=\frac{(2n+1)\pi}{\beta}\;;
\label{nd8}
\end{equation}
the functional integration becomes the determinant of the Green's function.

To evaluate the partition function with OPEP interaction ,  eqn.(\ref{nd5}),
an approximation scheme of boson loop expansion is used. This scheme has
been explained in details in the already quoted refs.
\cite{AlCeMoSa-88, CeSa-94, CeCoSa-97}. Basically, functional
integral over the fermion fields leaves an effective bosonic Lagrangian which
in the zeroth order gives rise to a mean field like that of RPA and in the
1st order to the one boson loop approximation.

This is achieved by availing of the Hubbard-Stratonovitch transformation
\cite{MoGaNaRa-74,Ke-70,Kl-78} to convert the potential part
inside the Feynman integrals  into a functional integral. For example,
the potential inside  eqn.(\ref{nd5}), under the
Hubbard-Stratonovitch transformation, becomes  
\begin{eqnarray}
\lefteqn{e^{-{i\over 2}\int_0^\beta\int d^3x \,d^3y 
\,\psi^\dagger(x)\Gamma_i\psi(x)
{\cal V}_\pi(|\bfm {\scriptstyle x}-\bfm
{\scriptstyle  y}|)\psi^\dagger(y)\Gamma_i\psi(y)}=\sqrt{\det (-{\cal V}_\pi)}}
\label{nd9}\\
&&\int D[\sigma_i]e^{{i\over 2}\int_0^\beta d\tau\int 
d^3x \,d^3y\sigma_i(x){\cal V}_\pi^{-1}(|\bfm {\scriptstyle x}-\bfm 
{\scriptstyle y}|)\sigma_i(y)
+i\int_0^\beta d\tau\int d^3x \sigma_i(x)\psi^\dagger(x)\Gamma_i\psi(x)},
\nonumber
\end{eqnarray}
where the measure over $\sigma_i$  is defined as
$$D[\sigma_i(x)]\longrightarrow\prod_{k}\frac{d\sigma_i(x_k)}{\sqrt{2\pi}}.$$
Performing the functional integration over the fermionic 
fields it becomes
\begin{equation}
Z=\sqrt{\det(-{\cal V}_\pi)}
\int D[\sigma_i]e^{\frac{1}{2}\sigma_i{\cal
V}_\pi\sigma_i}\det\left[{\cal G}_0^{-1}-\sigma_i\Gamma_i\right]\;.
\label{nd12}
\end{equation}
Since ${\cal G}_0^{-1}$ and $\sigma_i\Gamma_i$ commute :
\begin{equation}
Z=\sqrt{\det(-{\cal V}_\pi)}\det {\cal G}_0^{-1}
\int D[\sigma_i]e^{\frac{1}{2}\sigma_i{\cal
V}_\pi\sigma_i+\Tr\log\left[I-{\cal G}_0\sigma_i\Gamma_i\right]}.
\label{nd15}
\end{equation}

\section{Introduction of  $\Delta$'s, $\rho$-mesons and Short Range 
Correlations (SRC)}

The simplest way to introduce the $\Delta$'s in this formalism
is to generalise the nucleon field to a column vector 
\begin{equation}
\psi=\left({\psi_N\atop\psi_\Delta}\right)
\label{nd16}
\end{equation}
and the $\Gamma_i$ (eqn.\ref{nd1.3}) as
\begin{equation}
\Gamma_i=\left(\matrix{\bfm\sigma\cdot {\bf 
q}\tau_i&\frac{\displaystyle f_{\pi 
N\Delta}}{\displaystyle f_{\pi NN}}{\bf S}\cdot {\bf q}T_i\cr
\frac{\displaystyle f_{\pi N\Delta}}{\displaystyle 
f_{\pi NN}}{\bf S}^\dagger \cdot {\bf q}T_i^\dagger
&\frac{\displaystyle f_{\pi \Delta\Delta}}{\displaystyle 
f_{\pi NN}}\bfm{\cal S}\cdot {\bf q}
{\cal T}_i}\right)\;.
\label{nd16.1}
\end{equation}
Free ${\cal G}^0_\Delta$ becomes
\begin{equation}
{\cal G}^0_\Delta({\bf q},\omega_n)=-\frac{1}{\frac{q^2}{2M_\Delta}
+\delta M-\mu_\Delta-i\omega_n}
\label{nd19}
\end{equation}
where $\delta M=M_\Delta-M$. Matrices ${\cal S}_i$ and ${\cal T}_i$ are 
the generators of the 3/2 representation of the $SU(2)$ group ( see for
instance \cite{CeCoSa-97}).
 But $S_i$ and $T_i$ are the familiar 4 x 2
transition spin - isospin operators (see for example, \cite{BrWe-75}).

Further, a reasonable description of the dynamics of a nuclear system 
interacting via meson exchange needs to account, in order to be realistic,
of short range correlations (SRC). This SRC has been introduced in nuclear
systems in many ways, for example through the  Landau-Migdal parameter
$g^\prime$. It has been shown by \cite{CeCoSa-97} 
that at the one boson loop level,
beyond the Landau-Migdal theory, a momentum dependence of $g^\prime$
is necessary. Now there are  two independent parameters, namely the value
of the function $g^\prime(q)$ at the origin and a cut-off that kills the
effective interaction due to nucleon-nucleon (or nucleon-$\Delta$) short
range repulsion. Both parameters are density dependent, in a way
which is not known.

A simpler treatment to deal with SRC is given by Brown et al.
\cite{BrBaOsWe-77} which is successfully applied by Oset and coworkers  
(see, among many
other papers, \cite{CaOs-92}). It amounts to multiply the nucleon-nucleon
interaction ${\cal V}(r)$ by the two-body correlation function $g(r)$,
i.e.,
\begin{equation}
  \label{eq:bbow}
  {\cal V}(r)\longrightarrow g(r) {\cal V}(r)\;,
\end{equation}
and further to approximate $g(r)$ with
\begin{equation}
  \label{eq:j0}
  g(r)\simeq \frac{1}{3} j_0(q_c r)
\end{equation}
where $q_c$ is taken roughly equal to the $\omega$-mass.
This approach not only works well but also is almost parameter free.
Density dependence of the correlations is automatically taken care of.
It is to be noted that, it amounts to fix $g^\prime$ at about 0.7.

Of course a $\Delta$-hole pair can be excited by a $\rho$-meson as well.
We shall see in the following that a  $\rho$-meson exchange, with the
same scheme for SRC as before, can also be included in the formalism.
In the following calculations, $q_c^\pi = 770$ MeV/c and
$q_c^\rho = 1200$ MeV/c.

\section{The Loop Expansion}

We now expand the partition function (eq.\ref{nd15}) semi-classically. 
At the saddle point,i.e., at the mean field level, $\sigma_i \, = \, 0$ 
and the partition function becomes  
\begin{equation}
Z^{\rm mean ~field}=\sqrt{\det-{\cal V}_\pi}\det {\cal G}_0^{-1}\;.
\label{nd22}
\end{equation}

Quadratic deviations of the field from its mean field
value lead to 
\begin{equation}
\int D[\sigma_i]e^{\frac{1}{2}\sigma_i {\cal V}_\pi^{-1}\sigma_i-
\frac{1}{2}\Tr\sigma_i \Gamma_i {\cal G}_0 \sigma_j\Gamma_j{\cal G}_0}
=\left\{\det  {\cal V}_\pi^{-1}-\Tr \Gamma_i{\cal G}_0\Gamma_j{\cal G}_0
\right\}^{-\frac{1}{2}}\;
\label{nd22.2}
\end{equation}
so that the partition function (eq. \ref{nd15})  at the one loop order becomes
\begin{equation}
Z=\det{\cal G}_0^{-1}e^{Tr\log \left[I-{\cal V}_\pi
 \Gamma_i{\cal G}_0\Gamma_j{\cal G}_0\right]}=Z_0 e^{\Tr\log\left[ I-{\cal V}_\pi \Gamma_i{\cal G}_0\Gamma_j{\cal 
G}_0\right]}
\label{nd23}
\end{equation}
where
\begin{equation}
Z_0=\prod_{q}\left(1+e^{
-\beta\left[\frac{q^2}{2M}-\mu_N\right]}\right)
\prod_{q}\left(1+e^{
-\beta\left[\frac{q^2}{2M_\Delta}+\delta 
M-\mu_\Delta\right]}\right)\;.
\label{nd23.2.2}
\end{equation}

The grand potential, $\Omega=-\frac{1}{\beta}\,log Z $,  in one-loop order becomes  
\begin{equation}
\Omega^{\rm 1-loop}=-\frac{1}{\beta}\left\{\log Z_0
+\Tr\log\left[ I-{\cal V}_\pi
 \Gamma_i{\cal G}_0\Gamma_j{\cal G}_0\right]\right\}
\label{nd24}
\end{equation}
Thus for the 0-order part we get
\begin{equation}
\Omega_0=-\frac{V}{\beta}\intt{q}\left\{
\log\left[1+e^{-\beta\left(\frac{q^2}{2M}
-\mu_N\right)}\right]
+\log\left[1+e^{-\beta\left(\frac{q^2}{2M_\Delta}+\delta 
M-\mu_\Delta\right)}\right]
\right\}
\label{nd25}
\end{equation}
Derivatives of $\Omega $ with respect to $\mu _N$ and $\mu_\Delta$ result in
$N $ and $\Delta $ numbers respectively. Therefore,
\begin{eqnarray}
<N_N>&=&V\intt{q}n_N^0(q)
\label{n0n}\\
<N_\Delta>&=&V\intt{q}n_\Delta^0(q)
\label{n0d}
\end{eqnarray}
where
\begin{eqnarray}
n_N^0(q)&=&\frac{1}{1+e^{\beta\left(\frac{q^2}{2M}-\mu_N\right)}}\\
n_\Delta^0(q)&=&\frac{1}{1+e^{\beta\left(\frac{q^2}{2M_\Delta}+\delta 
M-\mu_\Delta\right)}}\;.
\end{eqnarray}

For  first order correction we define 
\begin{equation}
\Pi^0=\frac{1}{q^2}\Tr\Gamma_i{\cal G}_0\Gamma_i{\cal G}_0
\label{nd26}
\end{equation}
(the trace makes $i\not=j$  vanishing) which turns out to be
the generalisation of the Lindhard function to the finite 
temperature case. We have three different  $\Pi^0\, -s$, corresponding to 
the N particle - N hole, the $\Delta$ particle - N hole and the $\Delta$ particle -$\Delta$ hole 
propagators. The last one vanishes at the $T\to0$ limit.

\begin{eqnarray}
\Pi^0_{NN}({\bf q},\nu_n)&=&4\intt{k}\frac{n^0_N(k)-n^0_N(q+k)}{
i\nu_n+\epsilon(k)-\epsilon(q+k)}
\label{x1}\\
\Pi^0_{N\Delta}({\bf q},\nu_n)&=&\frac{16}{9}
\intt{k}\frac{n^0_N(k)-n^0_\Delta(q+k)}{
i\nu_n+\epsilon(k)-\epsilon_\Delta(q+k)+\mu_\Delta-\mu_N}\nonumber\\
&+&(N \leftrightarrow \Delta)
\label{x2}\\
\Pi^0_{\Delta\Delta}({\bf q},\nu_n)&=&16\intt{k}\frac{n^0_\Delta(k)
-n^0_\Delta(q+k)}{i\nu_n+\epsilon_\Delta(k)-\epsilon_\Delta(q+k)}
\label{x3}
\end{eqnarray}
where the numbers in front are contributions from the spin-isospin traces and 
\begin{equation}
\nu_n=\frac{2n\pi}{\beta}\;.
\label{nd27}
\end{equation}
The parity of these function with respect to $\nu_n$ simplify them further.

Thus the total $\Pi$ is given by
\begin{equation}
\Pi=f_{\pi N N}^2 \Pi^0_{NN}+f_{\pi N \Delta}^2 \Pi^0_{N\Delta}
+f_{\pi \Delta\Delta}^2 \Pi^0_{\Delta\Delta}\;.
\label{eqx1}
\end{equation}

so that one loop grand potential (eq.\ref{nd24}) is
\begin{equation}
\Omega^{\rm 1-loop}=\Omega_0-\frac{1}{\beta}\Tr\log\left[ I+{\cal V}_\pi q^2
\Pi \right]
=\Omega_0+\frac{1}{\beta}\sum_{n=1}^{\infty}
\frac{1}{n}\Tr\left[{\cal V}_\pi\Pi\right]^n
\label{nd29}
\end{equation}

Using the identity
$\frac{1}{n}=\int\limits_{0}^{1}\lambda^{n-1}d\lambda$ and  substituting ${\cal V}_\pi^\lambda=\lambda{\cal V}_\pi$ we get
\begin{equation}
\Omega^{\rm 1-loop}=\Omega_0+\frac{1}{\beta}\sum_{n=1}^{\infty}
\int\limits_{0}^{1}\frac{d\lambda}{\lambda}\Tr\left[{\cal 
V}_\pi^\lambda\Pi\right]^n\, .
\label{nd31}
\end{equation}
Further, summation of the series results in 
\begin{equation}
\Omega^{\rm 
1-loop}=\Omega_0+\frac{1}{\beta}\int\limits_{0}^{1}\frac{d\lambda}{\lambda}
\Tr\frac{{\cal V}_\pi^\lambda\Pi}{1-{\cal V}_\pi^\lambda\Pi}\, .
\label{nd32}
\end{equation}
The remaining trace means integration over the 3-momentum and sum over the
frequencies.

Number of $\Delta$'s and nucleons can now be obtained easily. 
\begin{equation}
<N_\Delta>=<N_\Delta>^{0}-\frac{1}{\beta}
\Tr{{\cal V}_\pi\over 1-{\cal V}_\pi \Pi}\frac{\partial\Pi}{\partial\mu_\Delta}
\label{nd34}
\end{equation}
and 
\begin{equation}
<N_N>=<N_N>^{0}-\frac{1}{\beta}
\Tr{{\cal V}_\pi\over 1-{\cal V}_\pi \Pi}
\frac{\partial\Pi}{\partial\mu_N}
\label{nd35}
\end{equation}
 
The 0-order quantities are  given by (\ref{n0n}) and (\ref{n0d}).
To evaluate eqs.(\ref{nd34}) and (\ref{nd35}) 
we find that 1st-order correction 
term (n=1 of eq.(\ref{nd29}))
$$<N_N>^{\rm 1st-ord}=-\frac{1}{\beta}\Tr\frac{\partial~}{\partial \mu_N}
\left[{\cal V}_\pi \Pi\right]\;$$ needs added attention because it is ill
defined as in the case of self-energy of the nucleon in a medium : 
\begin{equation}
\Sigma(k,\omega_n)=\frac{1}{\beta}\sum_{n^\prime}\intt{q}{\cal G}_0(
{\bf
q},\omega_{n^\prime}) {\cal V}_\pi(|{\bf k-q}|).
\label{eq2bb}
\end{equation}
We need a convergence factor $e^{i\omega_{n^\prime}\eta}$ which produces an
extra factor $n^0_N(q)$. The derivative with respect to $\mu_N$
gives

\begin{eqnarray}
\lefteqn{<N_N>^{\rm 1st-ord}=}
\label{1st}\\
&&12\intt{q}\intt{k}{\cal V}_\pi(|{\bf k-q}|)
\frac{\partial~}{\partial\mu_N}\left[n_N^0(q)n_N^0(k)\right]
\nonumber\\
&&+\frac{16}{3}\intt{q}\intt{k}\frac{\partial~}{\partial\mu_N}
\left[n_N^0(q)n_\Delta(k)
+n_N^0(k)n_\Delta(q)\right]{\cal V}_\pi(|{\bf k-q}|)
\nonumber
\end{eqnarray}
Similarly one gets for the $\Delta$ contribution
\begin{eqnarray}
\lefteqn{<N_\Delta^{\rm 1st-ord}>=}
\label{1st1}\\
&&\frac{16}{3}\intt{q}\intt{k}\frac{\partial~}{\partial\mu_N}
\left[n_N^0(q)n_\Delta(k)
+n_N^0(k)n_\Delta(q)\right]{\cal V}_\pi(|{\bf k-q}|)\nonumber\\
&&+48\intt{q}\intt{k}{\cal V}_\pi(|{\bf
k-q}|)\frac{\partial~}{\partial\mu_N}\left[n_\Delta(q)n_\Delta(k)\right]
\nonumber
\end{eqnarray}

The remaining part of the series can be evaluated straightforwardly :
\begin{equation}
-\frac{3}{\beta}\sum_{n}\intt{q}\frac{{\cal V}_\pi^2\Pi}{1-
{\cal V}_\pi\Pi}\frac{\partial\Pi}{\partial\mu_{N,\Delta}}\;.
\label{2nd}
\end{equation}
by evaluating $\Pi ^0$ -s (eqs. (\ref{x1}-\ref{x3}). 

Some remarks are not out of place here.
\begin{enumerate}
\item Sum in eqn. (\ref{2nd}) can be further transformed into an integral
over the imaginary axis. But this does not make numerical calculation easier.
\item the present formalism amounts to evaluate the diagram of fig.
\ref{fig:1}
\begin{figure}[d]
  \begin{center}
    \epsfig{file=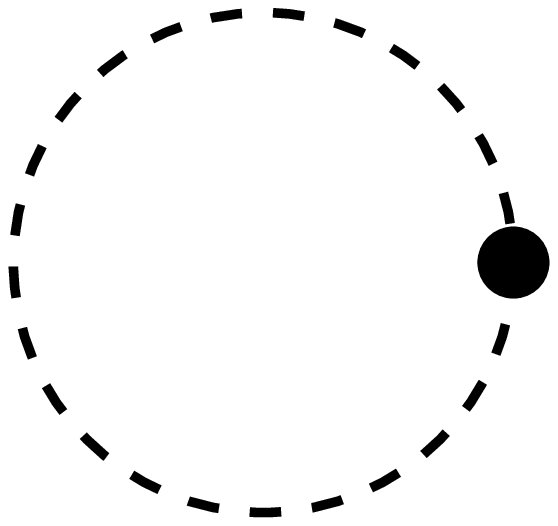,width=3cm,height=3cm}
    \caption{One-loop diagram for the number of $\Delta$ and nucleons.}
    \label{fig:1}
  \end{center}
\end{figure}
where the black dot denotes the $\partial\Pi/\partial\mu$ insertion and
the dashed circle the RPA-dressed pion. Since pions are spin-longitudinal
while the $\rho$'s are spin-transverse and in an infinite medium $\Pi$
conserves the elicity, the two kinds of mesons cannot mix together. Thus
the $\rho$ meson contribution simply amounts
to the similar diagram with the pion exchange potential
replaced by the $\rho$ exchange with  a factor 2 in front
of $\Pi$ coming from spin traces.
\item In general, $\partial\Pi\over\partial\mu$  in eqn. (\ref{2nd})
 obtained  from eqns. (\ref{nd26}-\ref{nd27})  becomes ill-defined for  $n=0$.
 But if  eqns. (\ref{nd26}-\ref{nd27}) describe
 analytical functions, even for $n=0$, the singularity can be 
separated from the integral and handled analytically.
\end{enumerate}

\section{Numerical results: densities and chemical potential}

The previous section provided the tool needed for the evaluation of
$<N_N>$ and $<N_\Delta>$. At the equilibrium $\mu_N=\mu_\Delta=\mu$
and hence both $<N_N>$ and $<N_\Delta>$ can be regarded as functions of $\mu$.
Putting for short $<N_N>=N_N(\mu)$ and $<N_\Delta>=N_\Delta(\mu)$
it is also clear that
\begin{equation}
  \label{eq:a11}
  N_N(\mu)+N_\Delta(\mu)=A=V\rho=V\frac{2k_F^3}{3\pi^2}\;,
\end{equation}
where $A$ is the baryon number and the last equality can be regarded as
a {\em definition} of an effective Fermi momentum. Here $\mu$ is the input of
the problem. If instead we choose $A$ in order to fix the thermodynamical 
conditions of the system we need to consider (\ref{eq:a11}) as an equation
for $\mu$ and solve it numerically.

Fig. \ref{fig:2}  displays the results of such a calculation.
It shows the  behaviour of the chemical potential at the equilibrium
for different values of the temperature and the
nuclear density. The graph contains only pion exchange at three
different levels of complexity, namely calculations up to second order
(dashed lines), full one-loop calculation (solid line) and full
calculation with dynamical pion (dotted line). We remember that in the
frame of Matsubara Green's function the pion propagator reads
$$\Delta_\pi={1\over \omega_n^2+{\bf q}^2+m_\pi^2}\;.$$
The figure
displays five bunches of lines, corresponding to five different
densities, namely (starting from the above), $\rho/\rho_0$ = 0.2, 0.4,
0.6, 0.8, 1; of course $\rho_0$ denotes the normal nuclear density.
Remarkably these three different dynamics are almost completely superimposed:
only looking very carefully at the figure one sees that the dashed line
is a little bit higher than the other two, that instead are 
practically coincident.
\begin{figure}[d]
  \begin{center}
    \epsfig{file=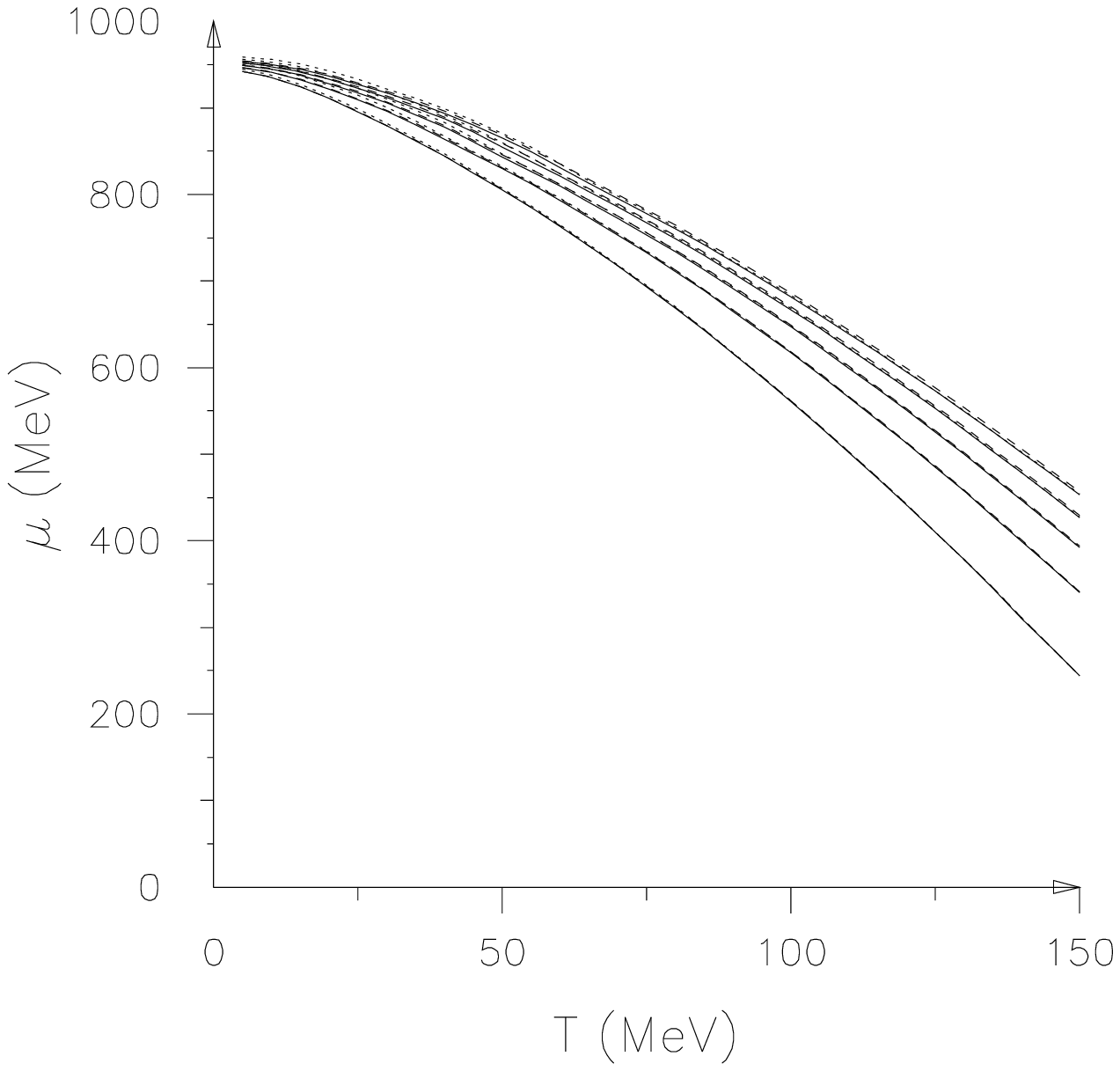,width=12cm, height=15cm}
    \caption{Chemical potential of a system of nucleons and $\Delta$'s 
      interacting via pion exchange as a function of the temperature.
      Lines as explained in the text.  }
    \label{fig:2}
  \end{center}
\end{figure}

The second plot we present (fig. \ref{fig:3}) displays instead the
ratio between nucleon and $\Delta$ density keeping fixed the sum
$\rho_N+\rho_\Delta=\rho$ again as a function of the temperature and
with five bunches as before. Of course this time the highest bunch
corresponds to the highest density.
\begin{figure}[d]
\begin{center}
\epsfig{file=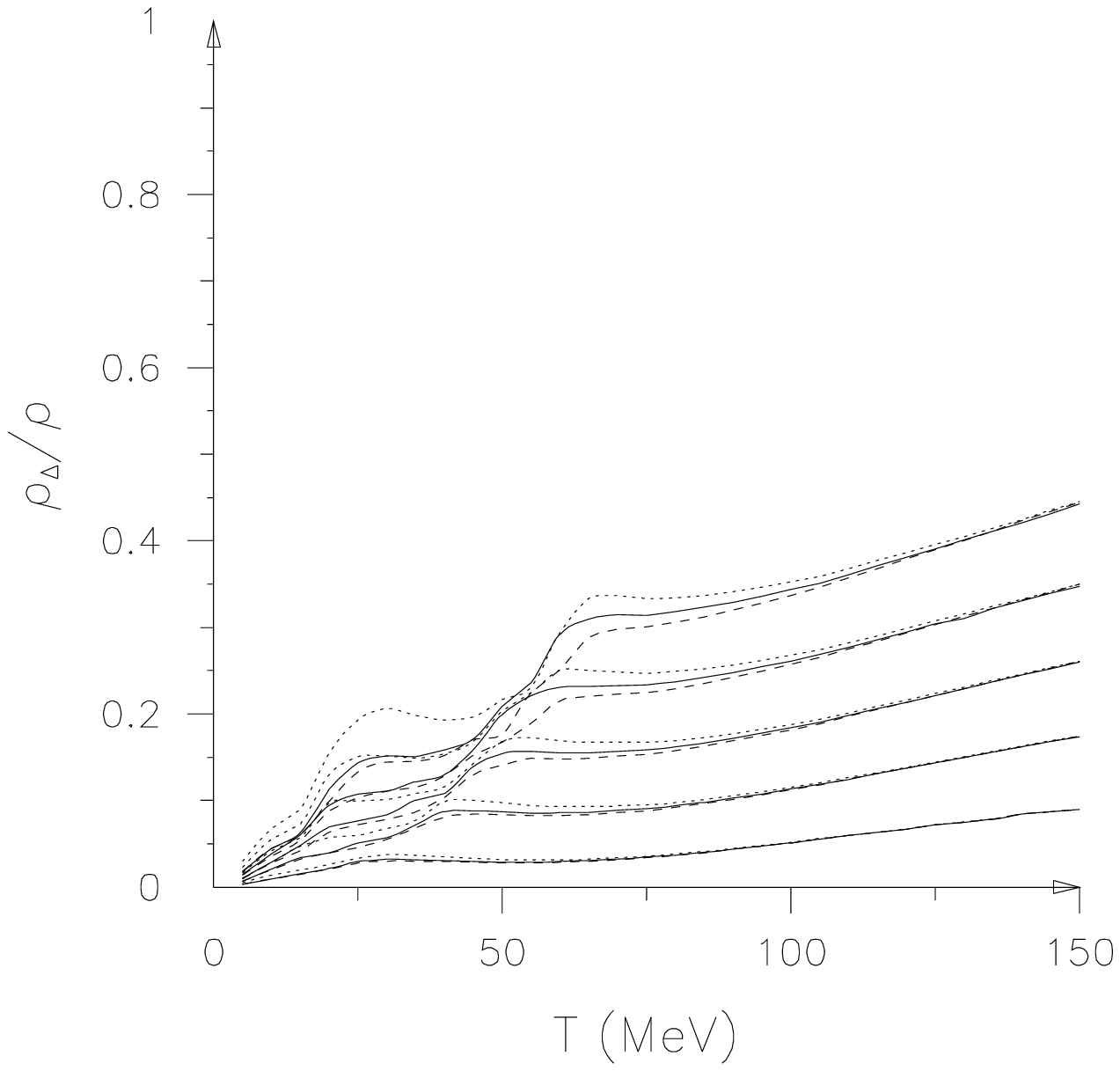,width=12cm, height=15cm}
  \caption{Ratio $\rho_\Delta/\rho_N$ as a function of the temperature. 
Lines as above. The total densities are 0.2 $\rho_0$, 0.4 $\rho_0$, 0.6 
$\rho_0$, 0.8 $\rho_0$ and 1 $\rho_0$ in ascending order. 
Only pions are accounted for.}
  \label{fig:3}
\end{center}
\end{figure}
Here the plots are much more sensitive to the dynamics.

Insofar we have only considered exchange of pions. Let us add some more 
dynamics by including the exchange of correlated $\rho$-mesons
the results are shown in figs. \ref{fig:4} and \ref{fig:5}.
\begin{figure}[d]
  \begin{center}
      \epsfig{file=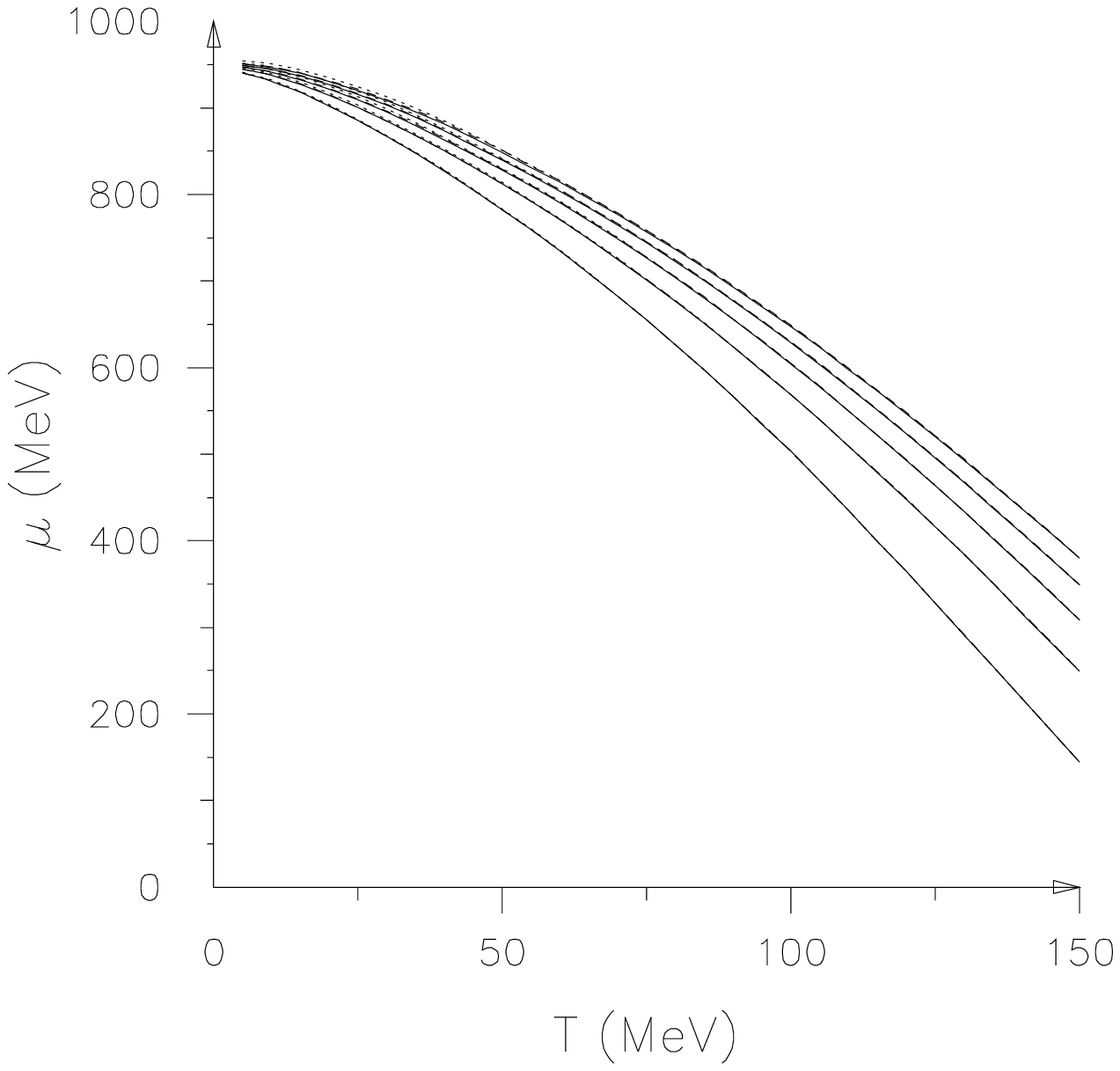,width=12cm, height=15cm}
    \caption{Chemical potential of a system of nucleons and $\Delta$'s 
      interacting via pion plus $\rho$ as a function of the
      temperature.  Lines as explained in the text.}
    \label{fig:4}
  \end{center}
\end{figure}
\begin{figure}[d]
\begin{center}
  \epsfig{file=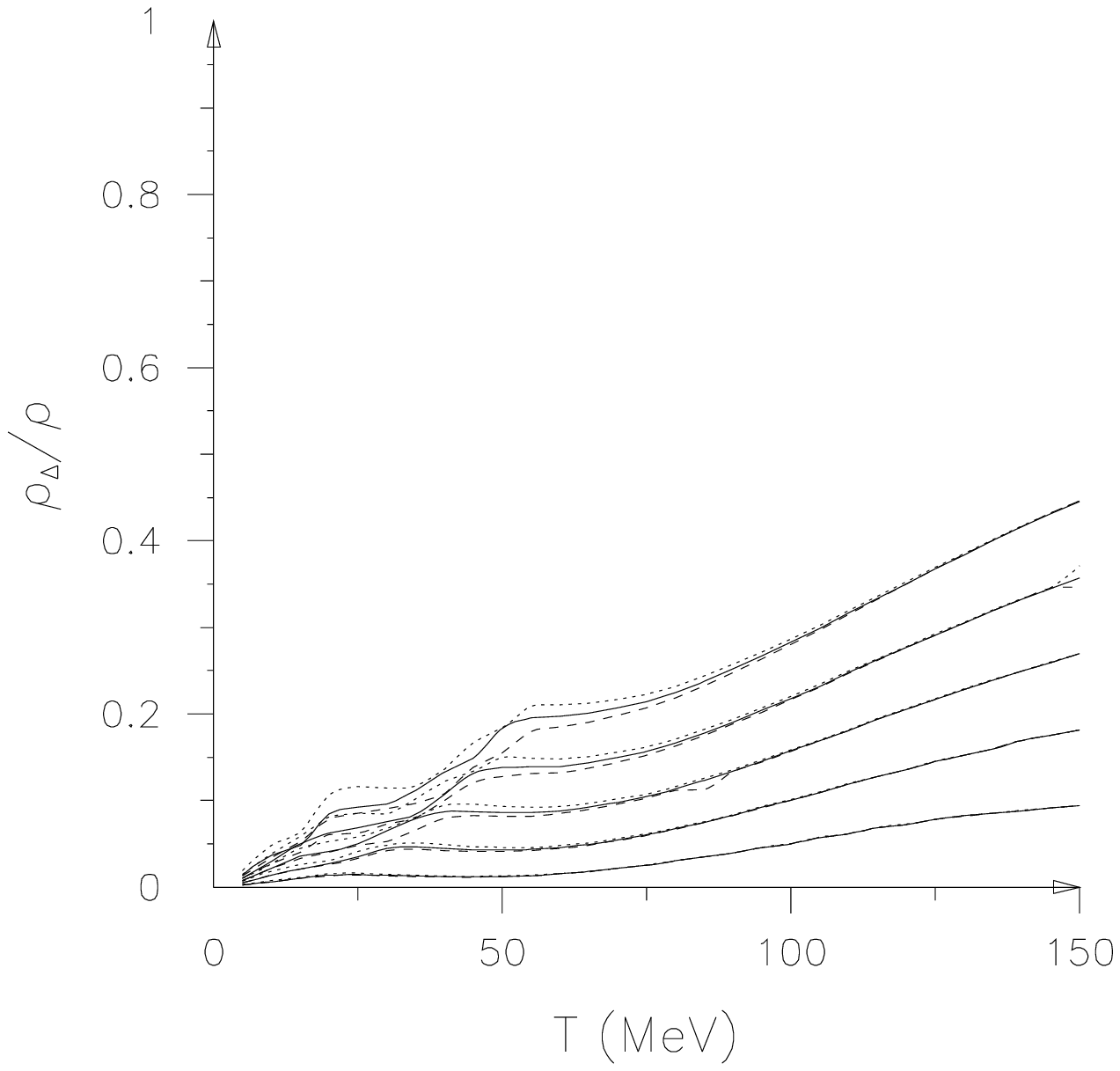,width=12cm, height=15cm}
  \caption{Ratio $\rho_\Delta/\rho_N$ as a function of the temperature, with 
pion plus $\rho$ exchange. Lines as in \protect\ref{fig:4}}
  \label{fig:5}
\end{center}
\end{figure}
Remarkably, the $\rho$-meson exchange lowers significantly the chemical 
potential, while the ratio $\rho_\Delta/\rho_N$ as a function
of the temperature is  sensible to it only in the low $T$ region, where
the full dynamical calculation shows a remarkable reduction of the
$\rho_\Delta/\rho_N$ ratio. At higher $T$ instead the effect fades out.
 
The model at hand contains only a few number of parameters. Coupling constants
and cut-off are more or less well defined (we use here standard coupling 
$C_\rho=2.3$) the only crucial parameter being the one connected to SRC, 
and in particular $q_c^\pi$. Thus in figs. \ref{fig:6} and \ref{fig:7}
we plot again the temperature and the ratio $\rho_\Delta/\rho_N$
for the most complete dynamical case ($\pi$ plus $\rho$ fully dynamic
at the one-loop order but with different values of $q_c^\pi$, namely
$q_c^\pi = 650 $ MeV/c, (solid line), $q_c^\pi = 770  $ MeV/c (dashed line)
 and $q_c^\pi = 900 $ MeV/c (dotted line); the same conventions as in the 
two previous figures are adopted.
\begin{figure}[d]
  \begin{center}
      \epsfig{file=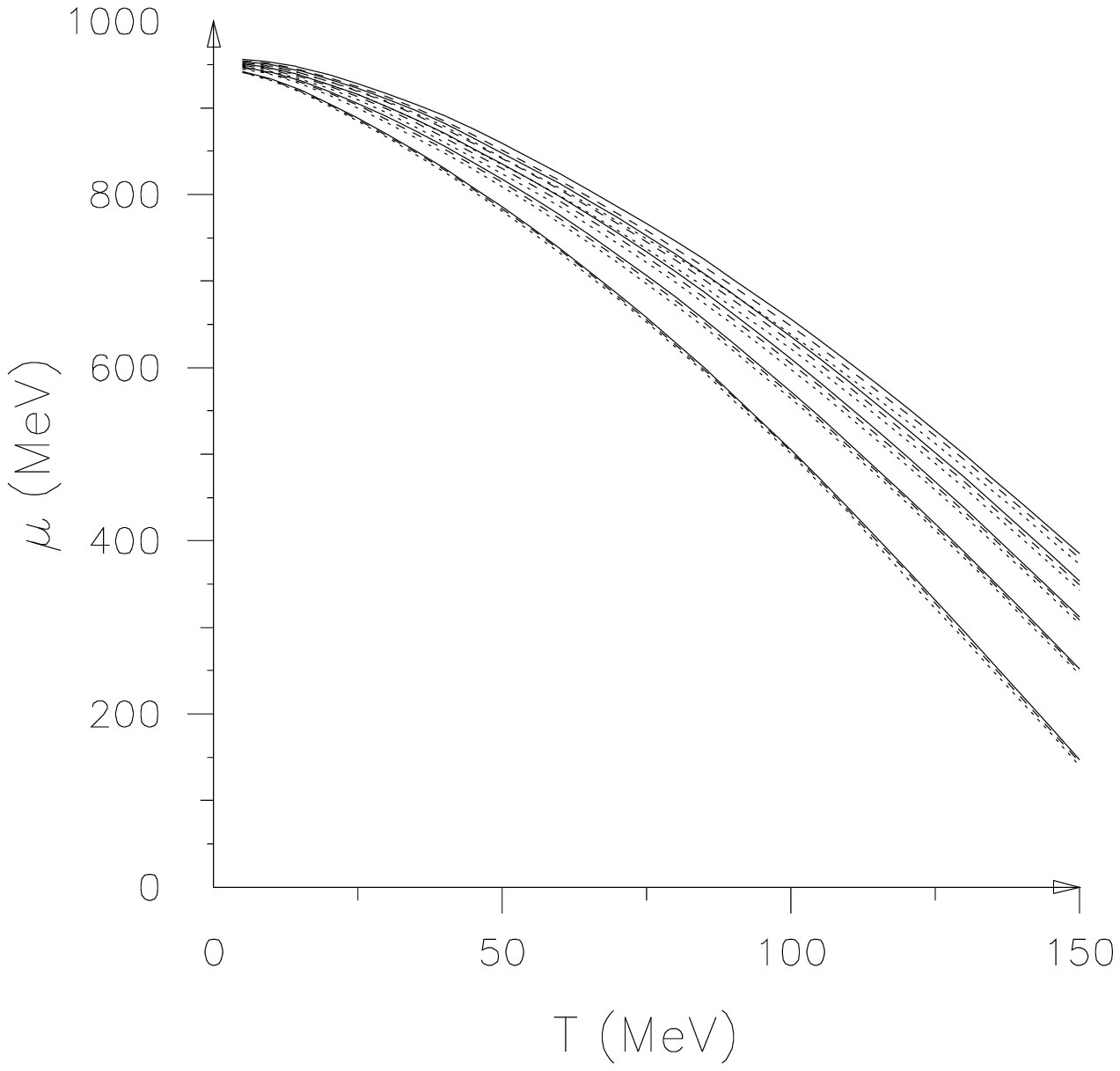,width=12cm, height=15cm}
    \caption{Chemical potential of a system of nucleons and $\Delta$'s 
      interacting via pion plus $\rho$ as a function of the
      temperature.  Lines as explained in the text.}
    \label{fig:6}
  \end{center}
\end{figure}
\begin{figure}[d]
\begin{center}
  \epsfig{file=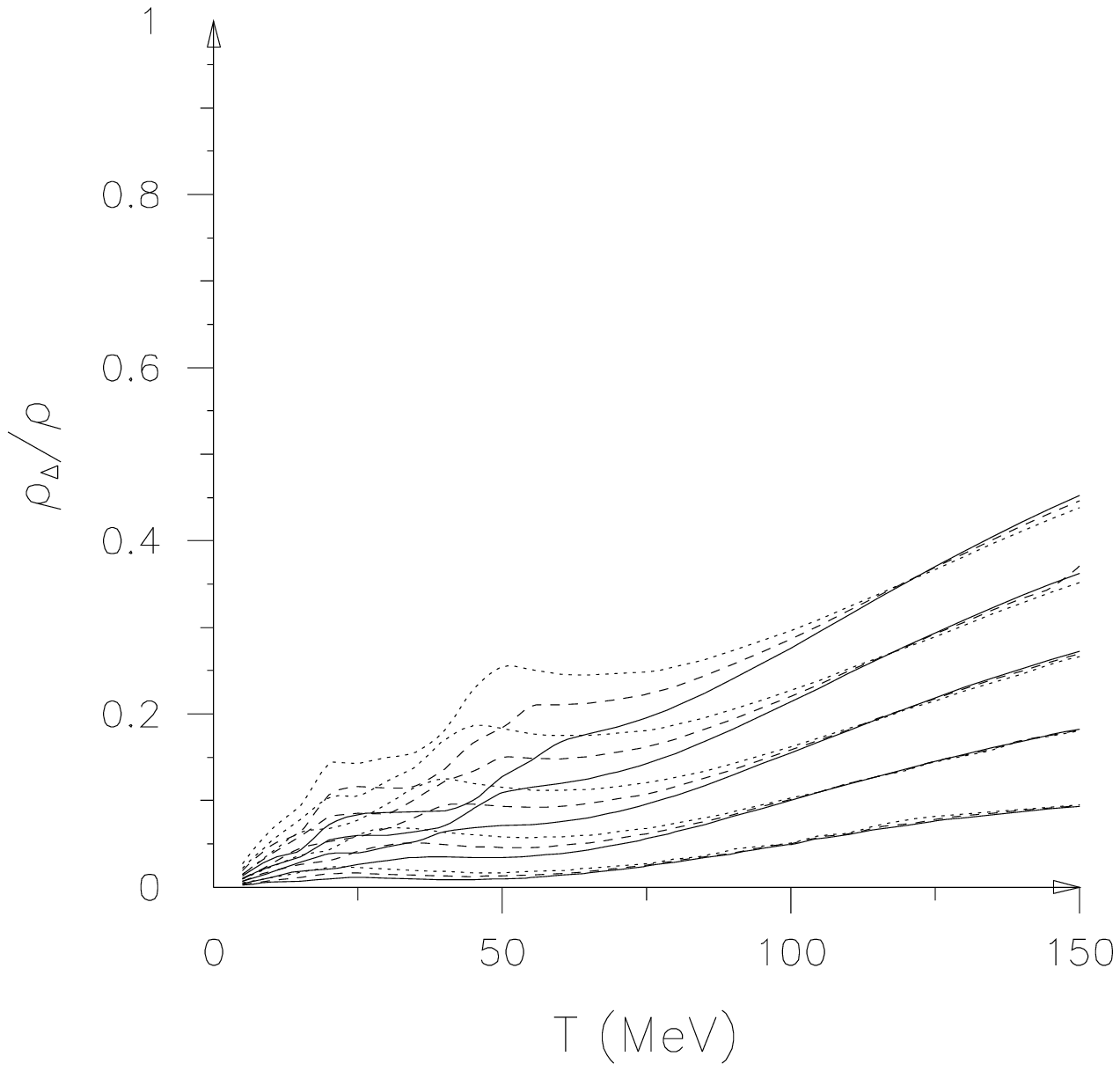,width=12cm, height=15cm}
  \caption{Ratio $\rho_\Delta/\rho_N$ as a function of the temperature, with 
pion plus $\rho$ exchange. Lines as in \protect\ref{fig:6}}
  \label{fig:7}
\end{center}
\end{figure}
We observe that while the chemical potential is almost insensitive to 
the variations of $q_c^\pi$ the ratio  $\rho_\Delta/\rho_N$, on the 
contrary, shows that this ratio sensibly increases with  $q_c^\pi$, 
in particular in an intermediate range of the temperature, 
while at temperatures of the order of 150 MeV the effect seems to fade out.

\section{Numerical results: the momentum distribution}

The momentum distribution displays an impressive dependence upon the 
temperature. We define the  momentum distribution through the expression
\begin{eqnarray}
  \label{eq:momdist}
  <N>_N&=&\Tr \int\frac{d^3 q}{(2\pi)^3}n_N(q)\\
  <N>_\Delta&=&\Tr \int\frac{d^3 q}{(2\pi)^3}n_\Delta(q)
\end{eqnarray}
the trace being taken over spin-isospin degrees of freedom 
(4 for nucleons, 16 for $\Delta$'s).
In practice this amounts to take the integrand of all the previous 
calculations. Thus the result is a byproduct of the previous calculation.
The results are plotted in fig. \ref{fig:8} for different 
temperatures and densities. The normalisation is everywhere to
a $\theta(k_F-q)$ when $T\to0$.
\begin{figure}[d]
\begin{center}
\mbox{
  \begin{tabular}{ccc}
\epsfig{file=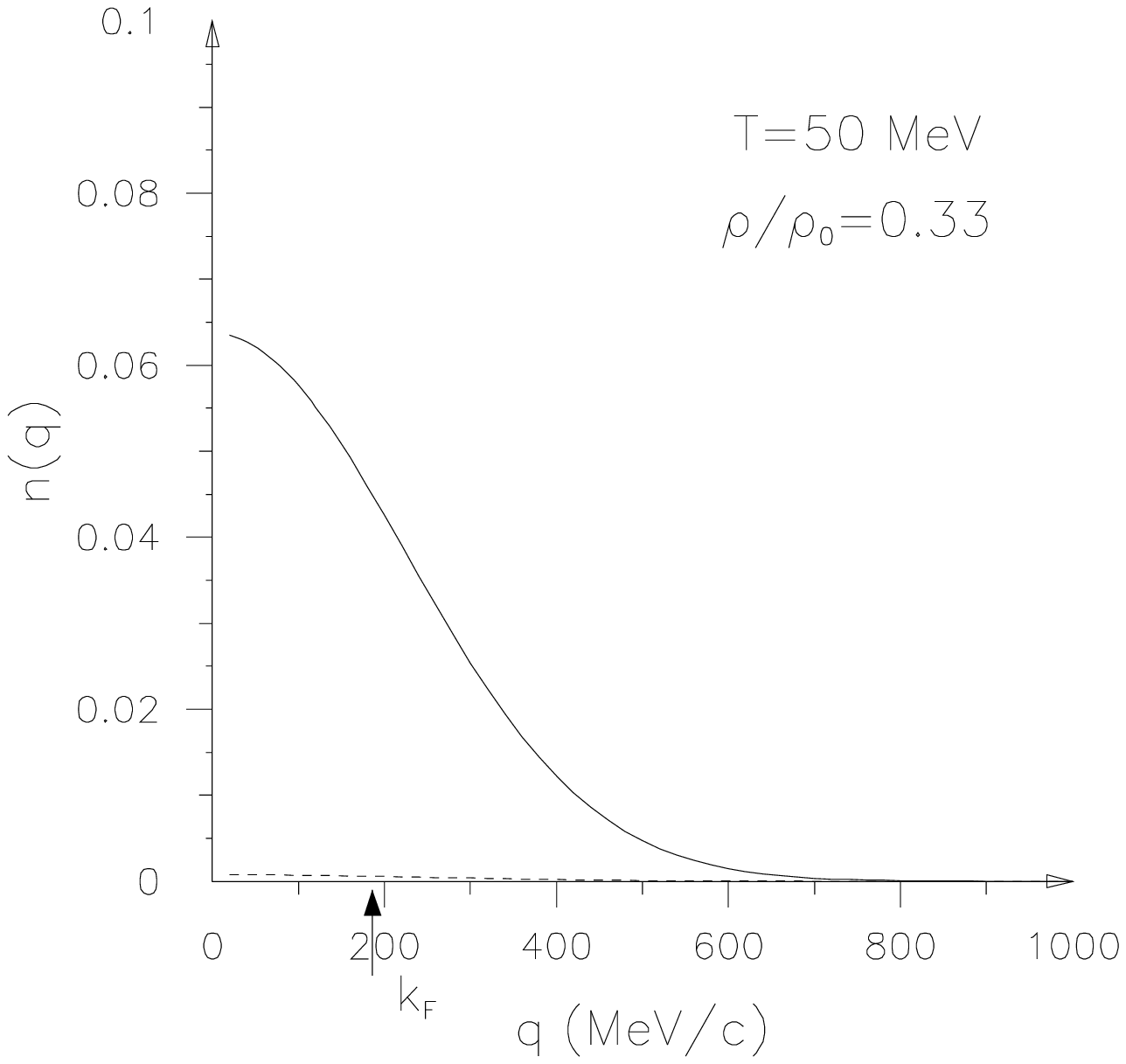,width=4.1cm,height=5.5cm}
&
\epsfig{file=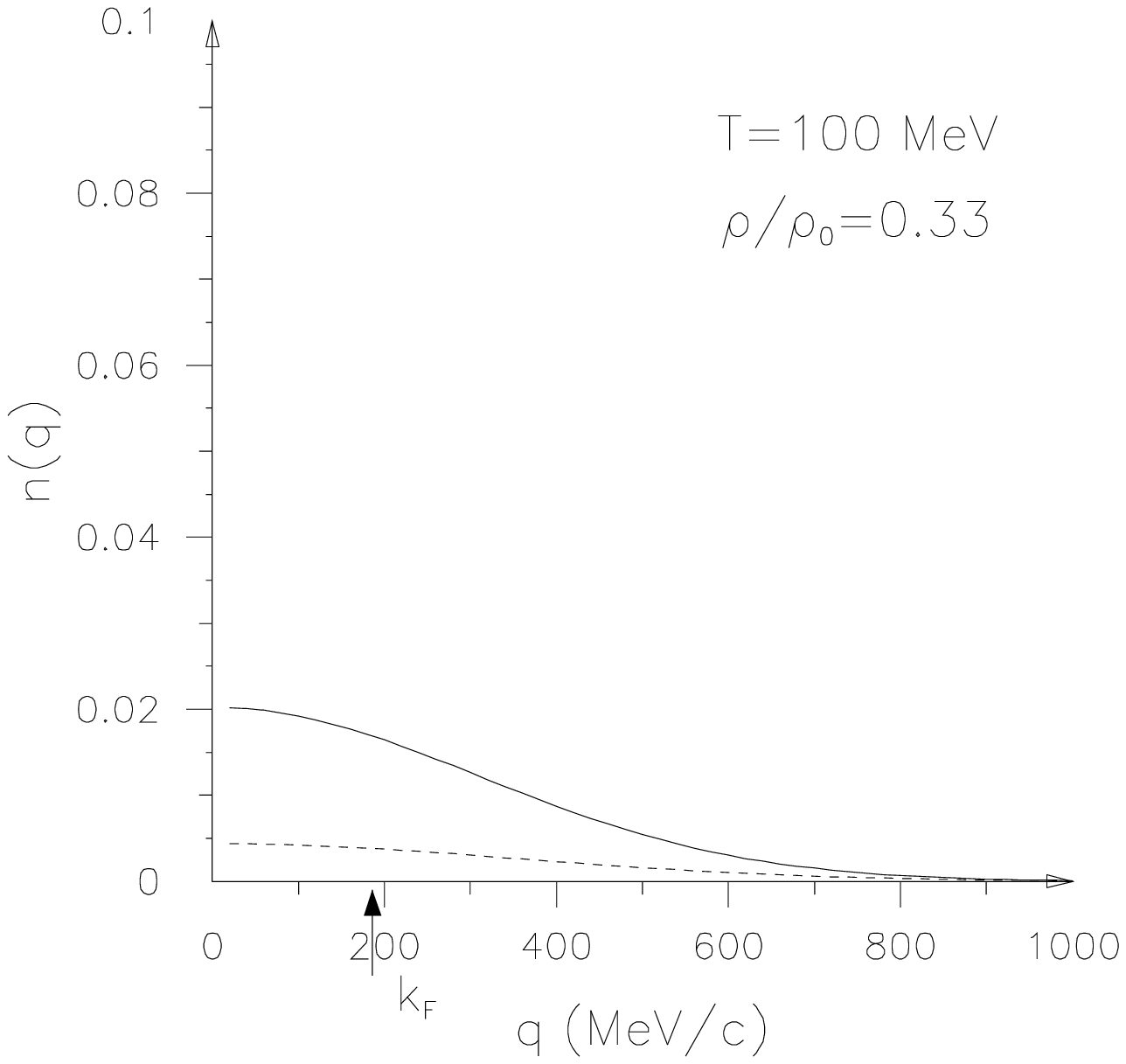,width=4.1cm,height=5.5cm}
&
\epsfig{file=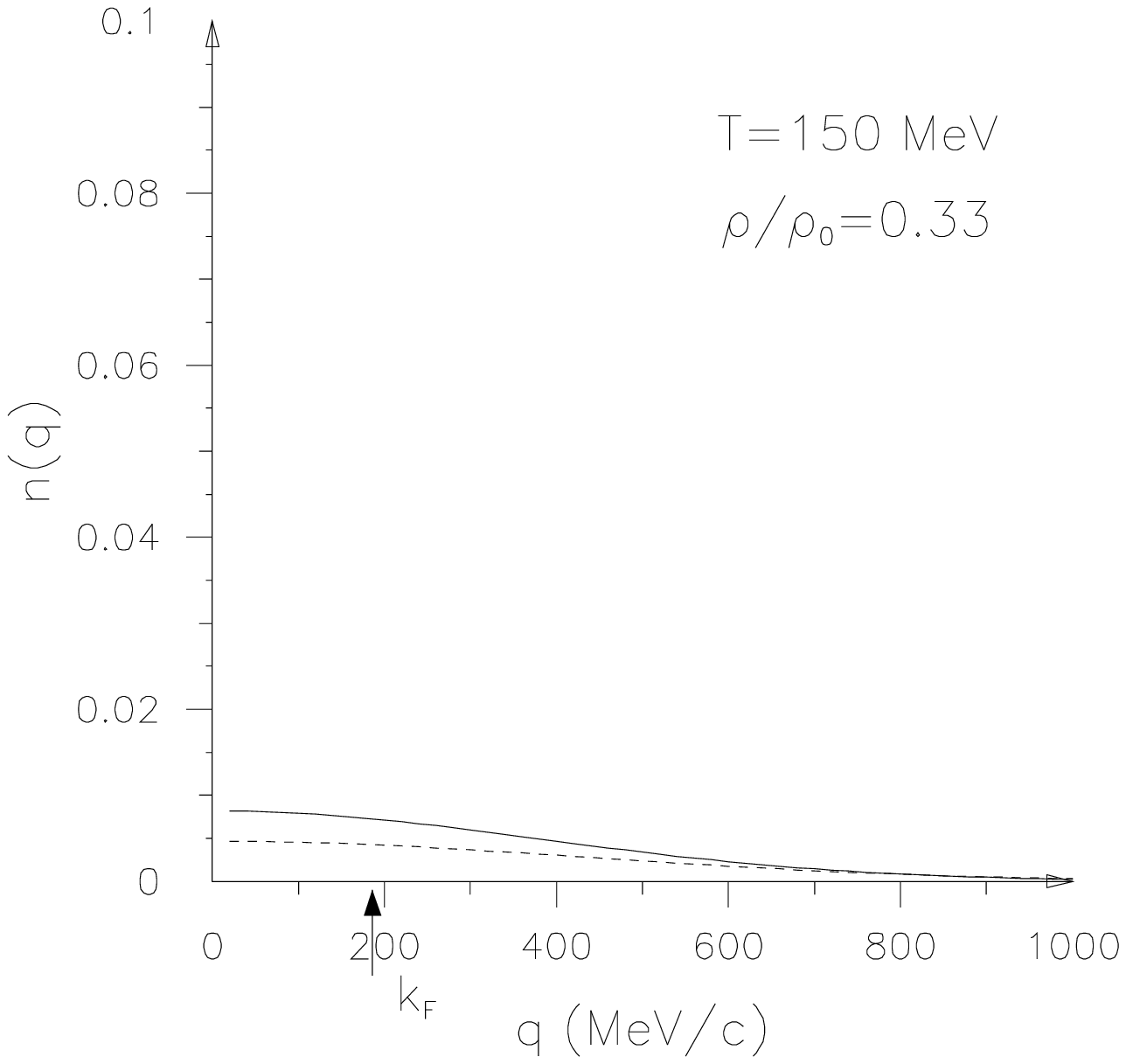,width=4.1cm,height=5.5cm}
\\
\epsfig{file=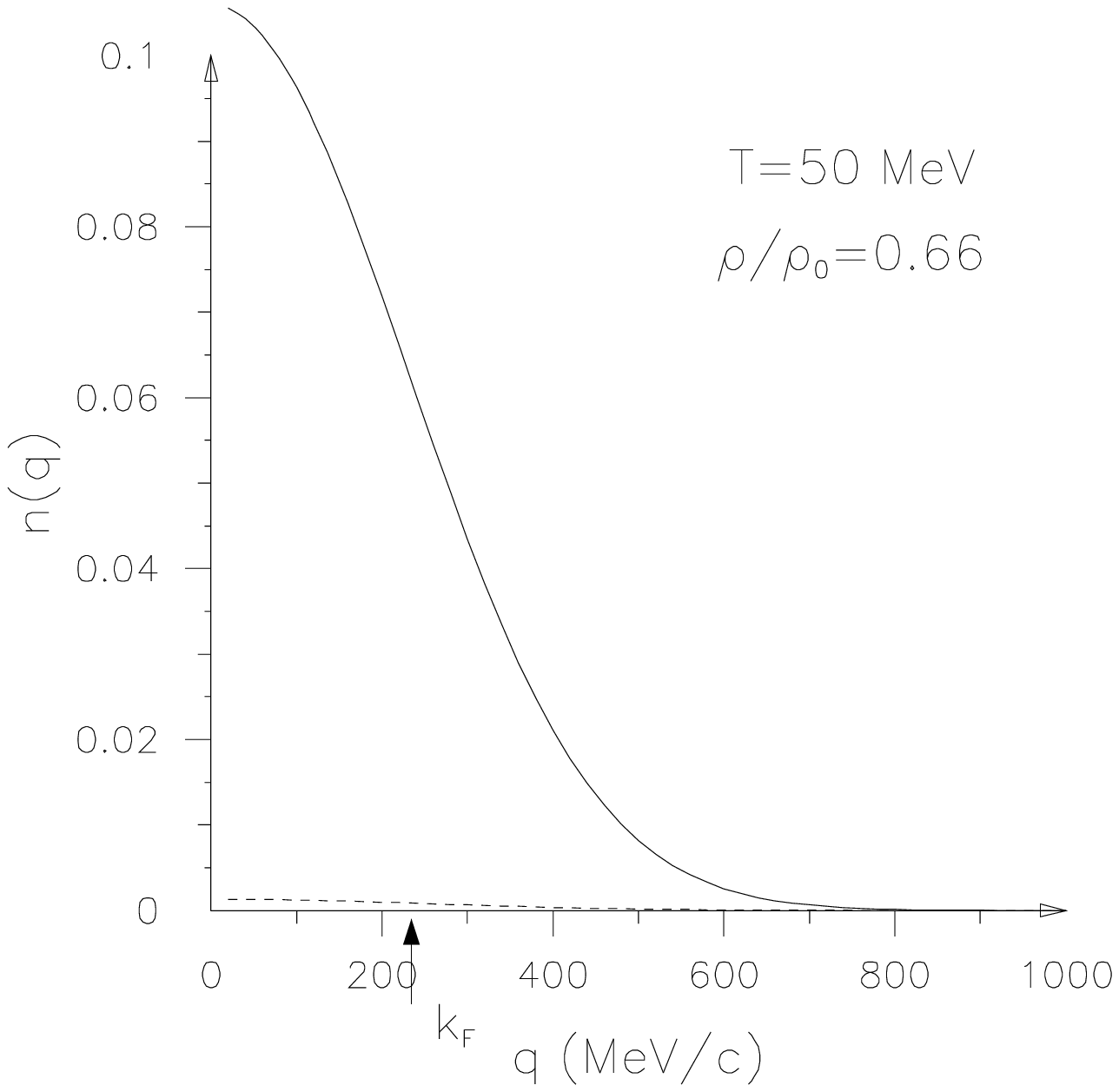,width=4.1cm,height=5.5cm}
&
\epsfig{file=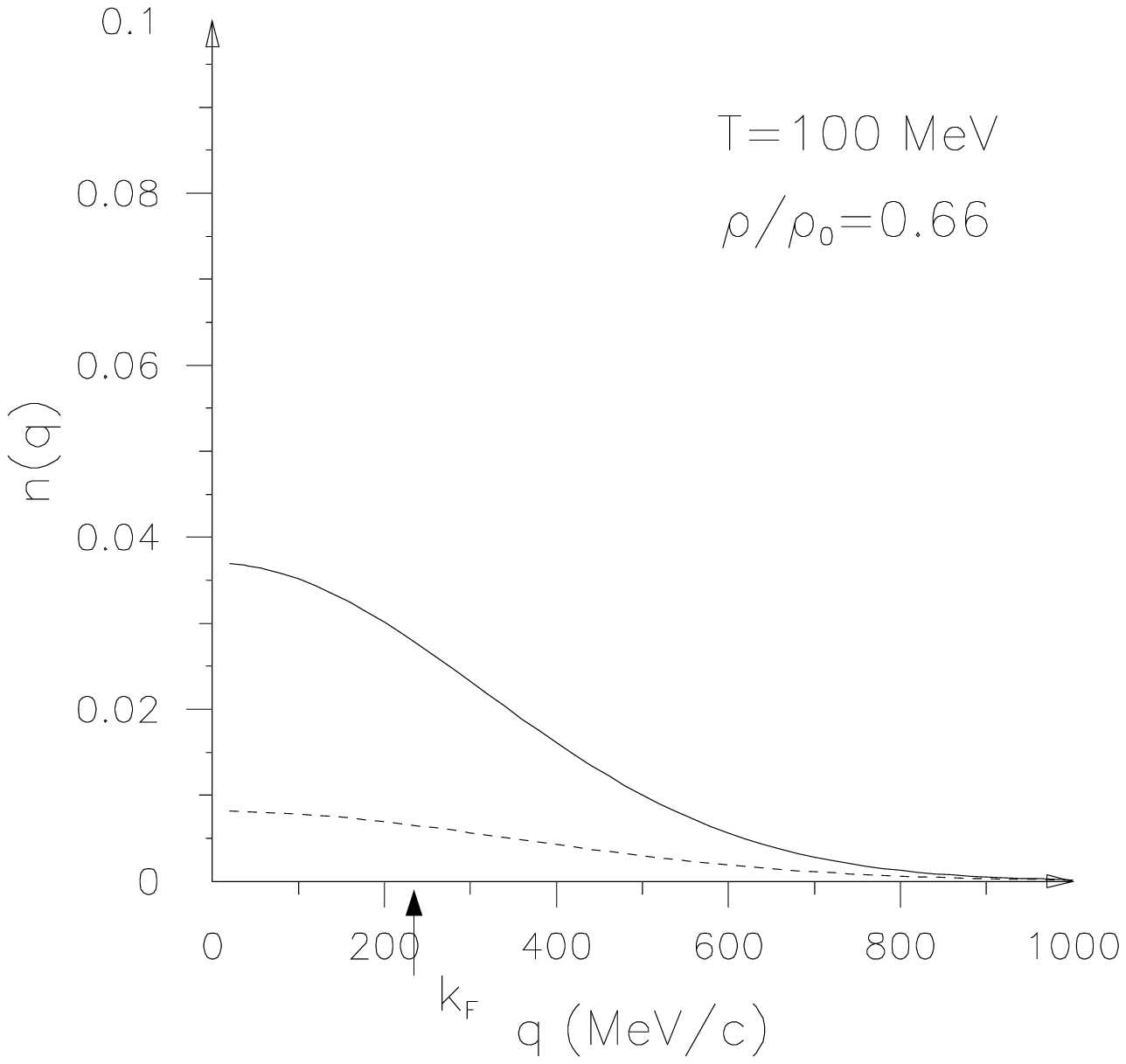,width=4.1cm,height=5.5cm}
&
\epsfig{file=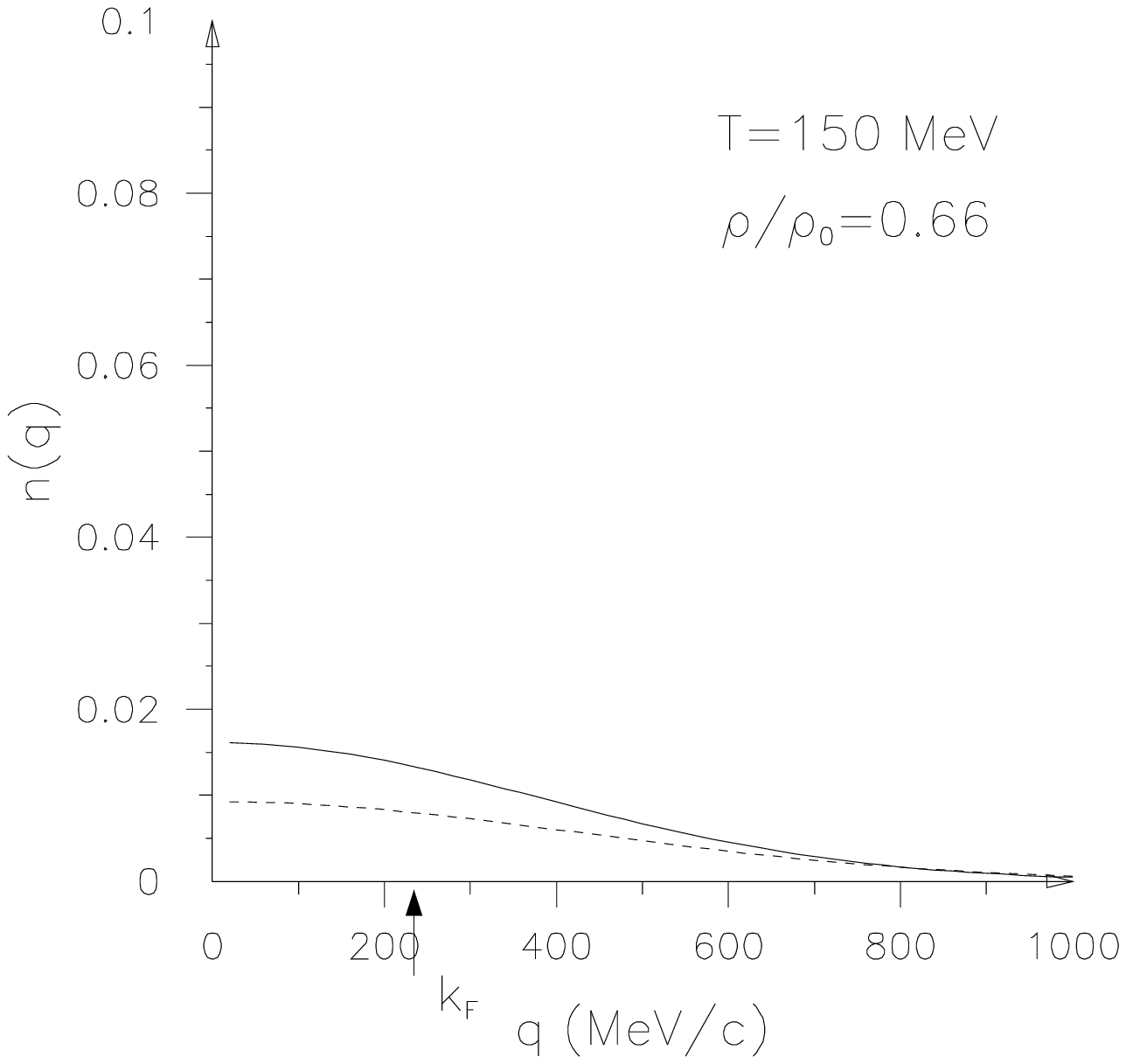,width=4.1cm,height=5.5cm}
\\
\epsfig{file=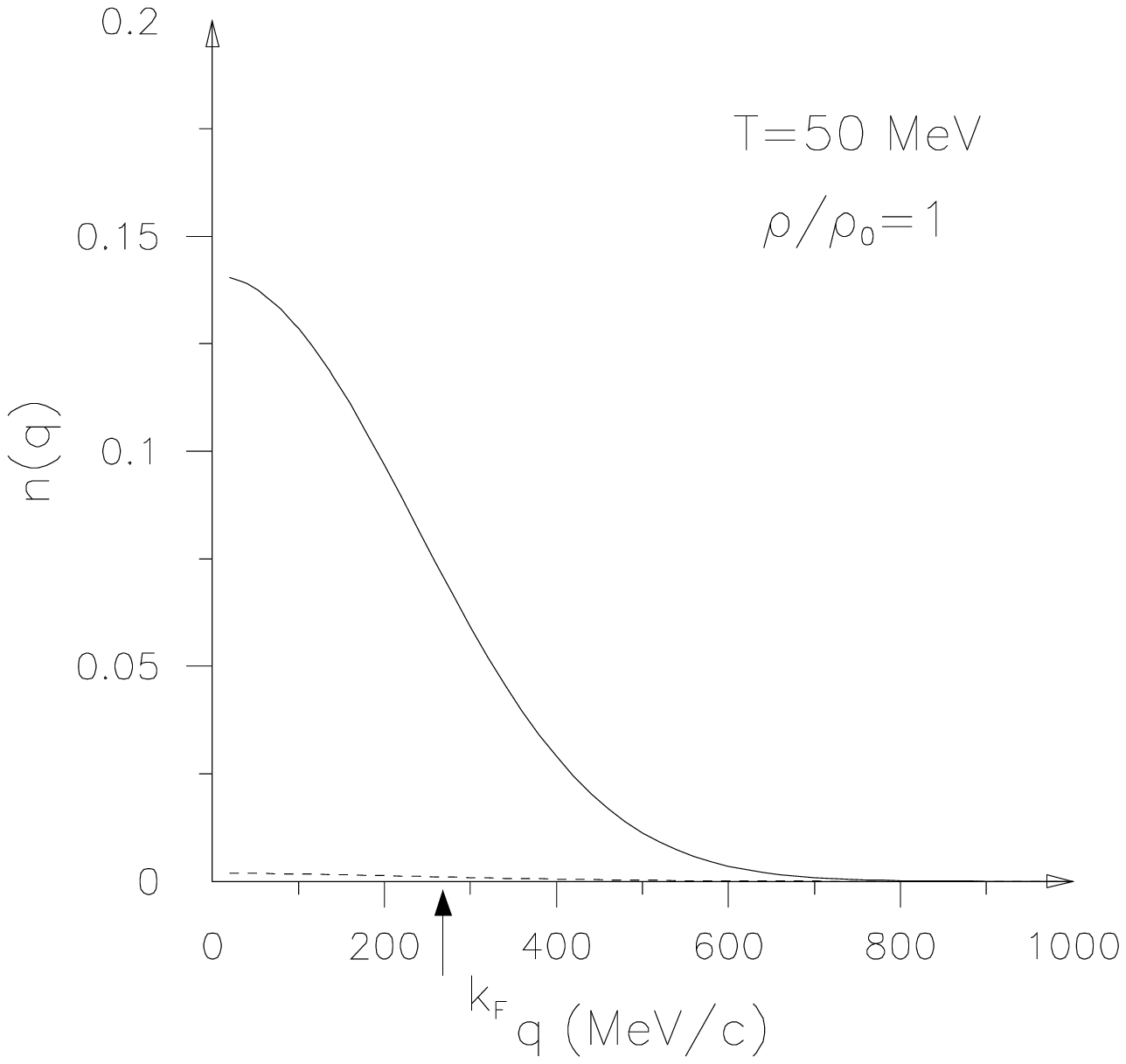,width=4.1cm,height=5.5cm}
&
\epsfig{file=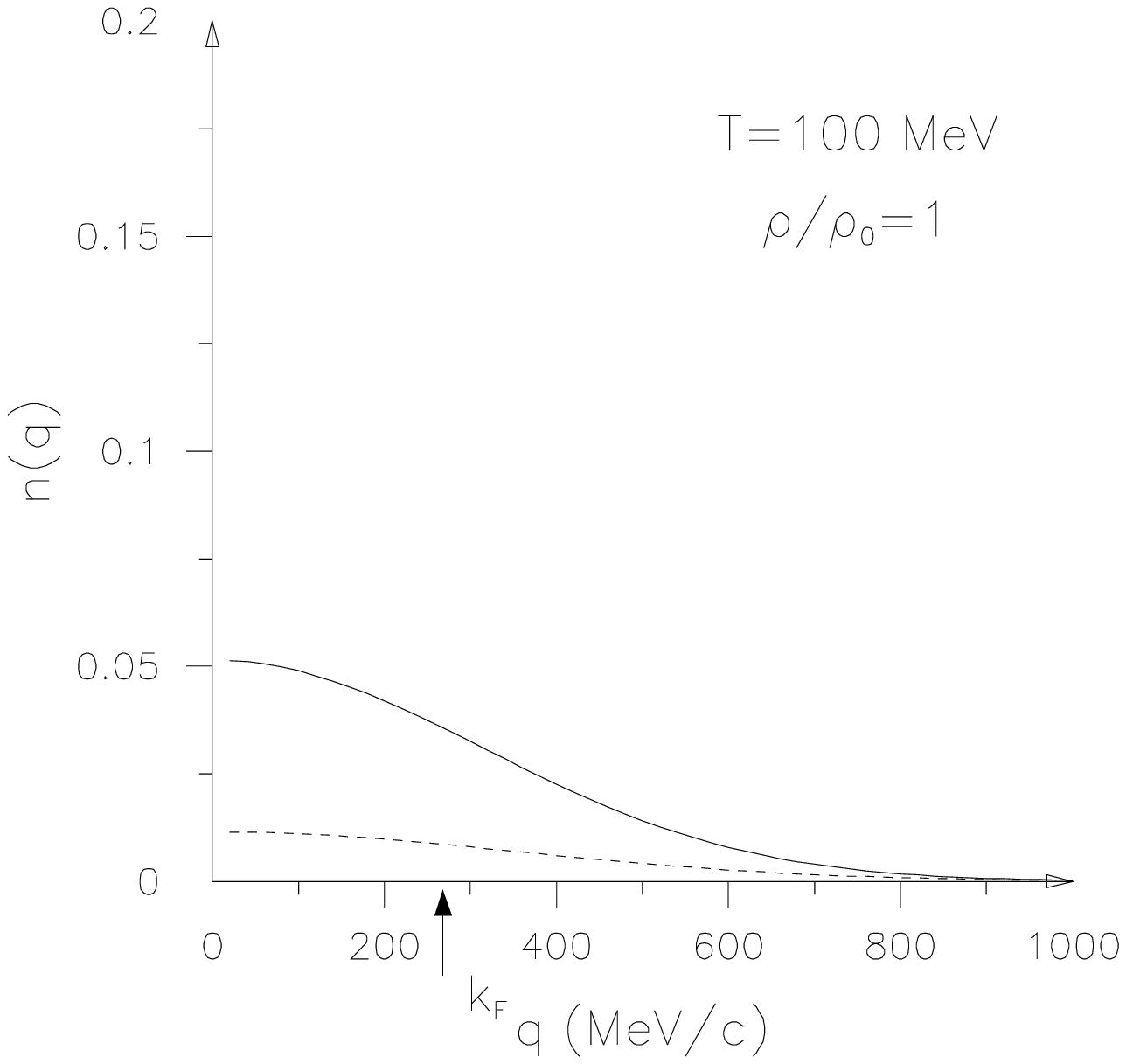,width=4.1cm,height=5.5cm}
&
\epsfig{file=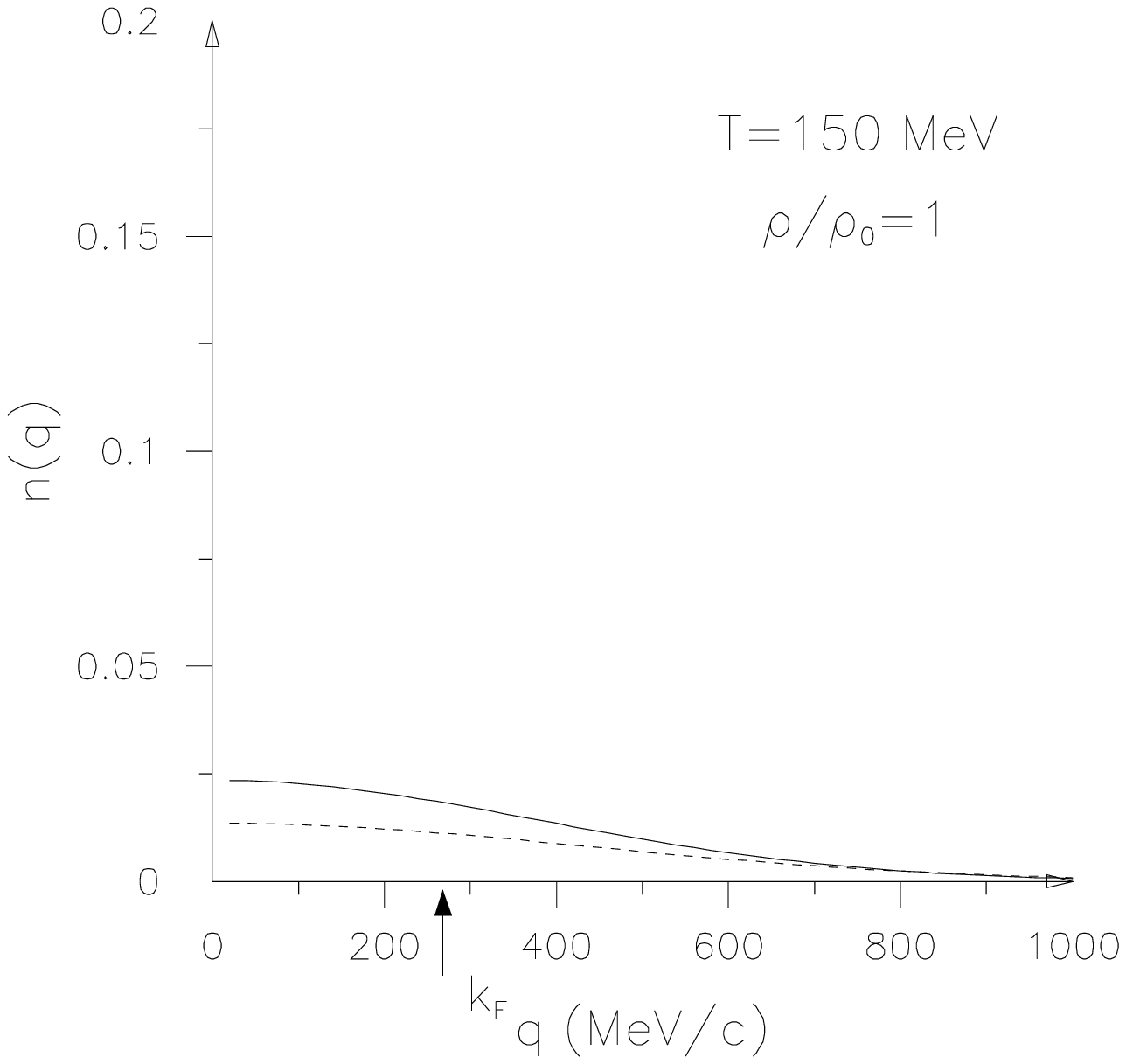,width=4.1cm,height=5.5cm}
  \end{tabular}}
  \caption{Momentum distribution of nucleons (solid line) and $\Delta$'s
(dashed line) at different densities and temperatures}
  \label{fig:8}
\end{center}
\end{figure}

It is impressive to remark that the distribution is immediately spreaded
to high momenta with the increasing of the temperature. Remind that 
the distribution is weighted with a $q^2$ in the Jacobian, so that 
the high momentum component become more and more important. An even more
striking insight can be obtained by plotting the quantity $q^2n(q)$
and by comparing it with the Free Fermi Gas result at 0 temperature, as we do
in fig \ref{fig:9}. There it is clear that the areas of the curves is preserved
while the shape is enormously modified.

\begin{figure}[d]
\begin{center}
\mbox{
  \begin{tabular}{ccc}
\epsfig{file=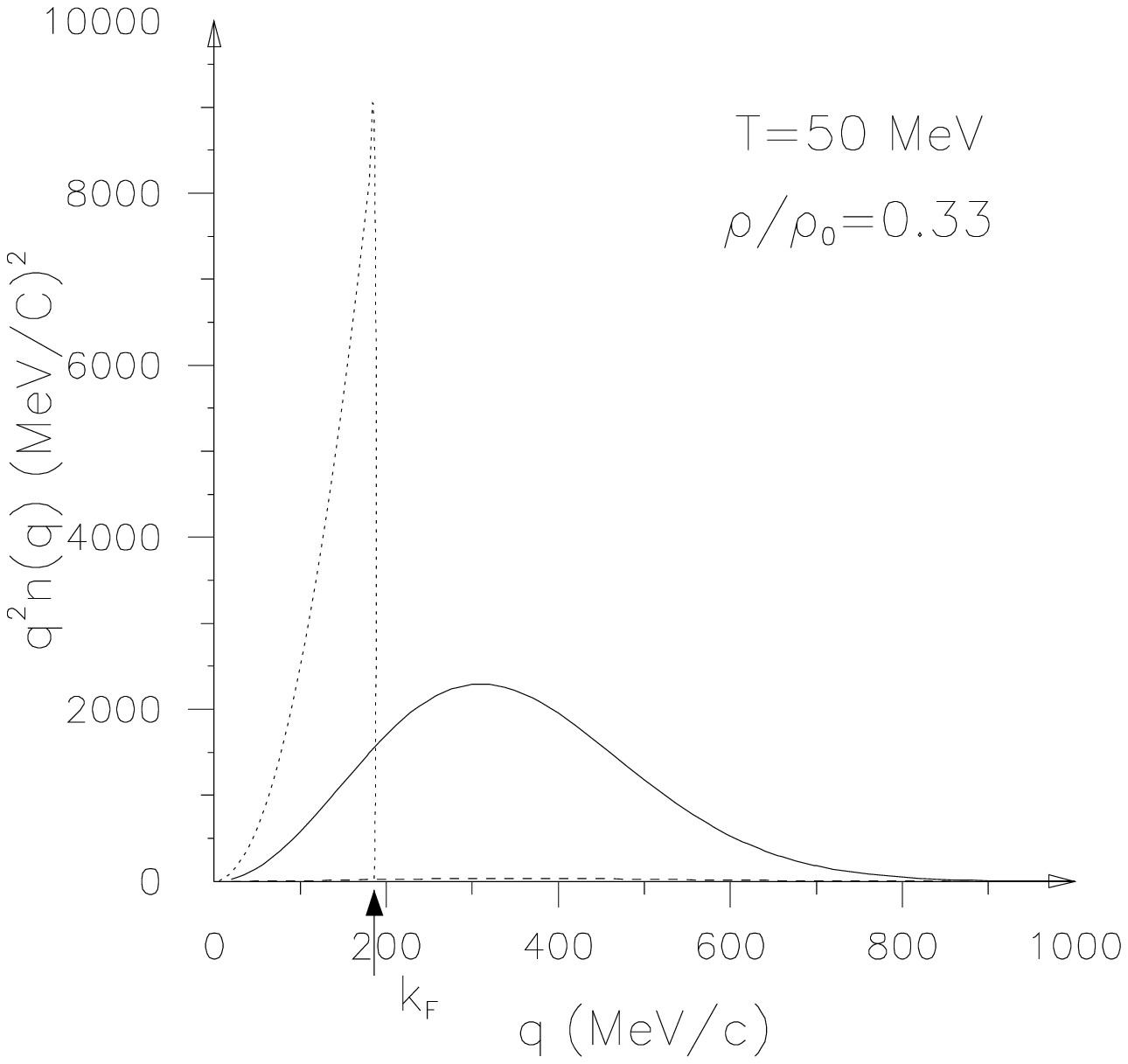,width=4.1cm,height=5.5cm}
&
\epsfig{file=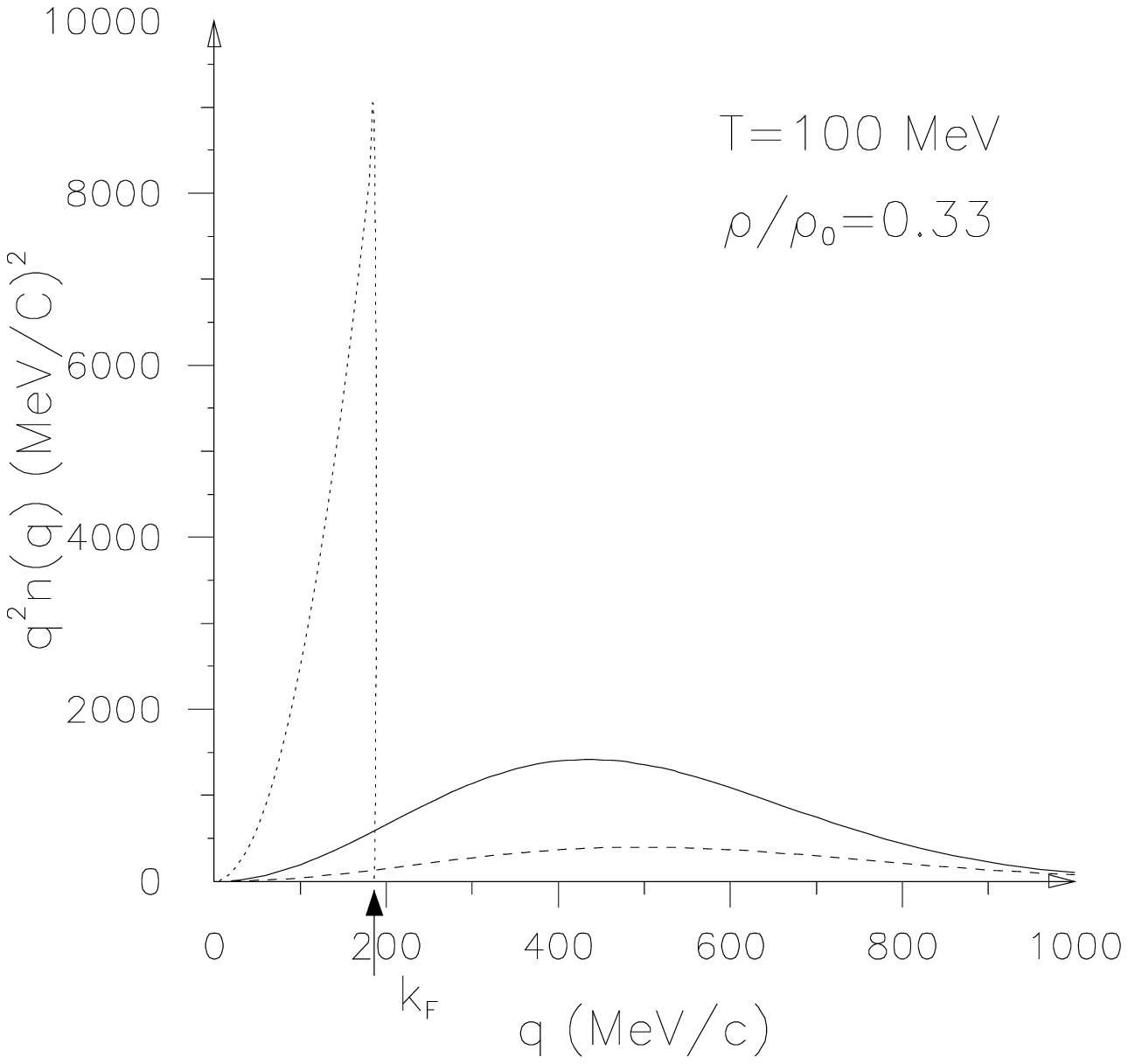,width=4.1cm,height=5.5cm}
&
\epsfig{file=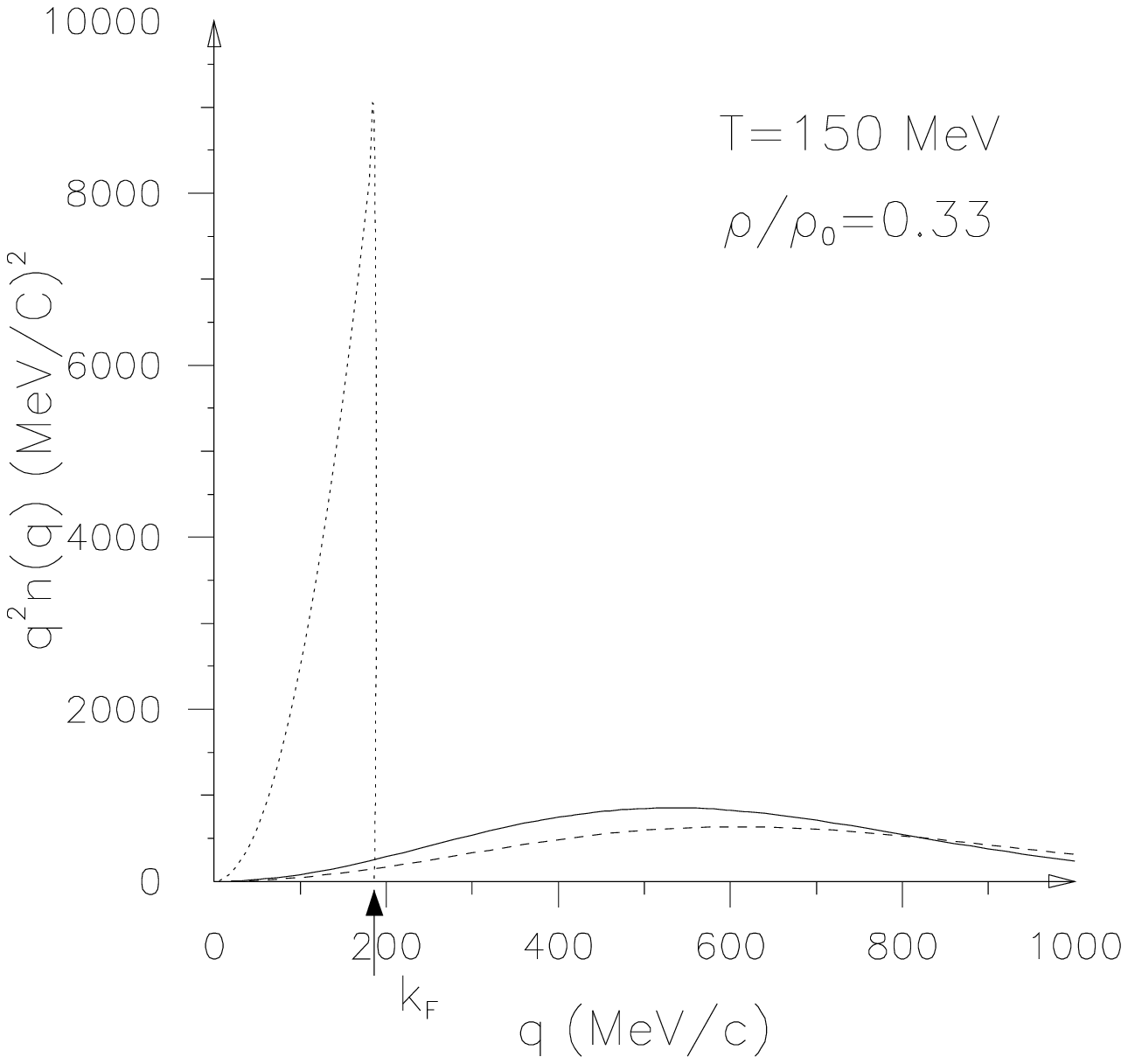,width=4.1cm,height=5.5cm}
\\
\epsfig{file=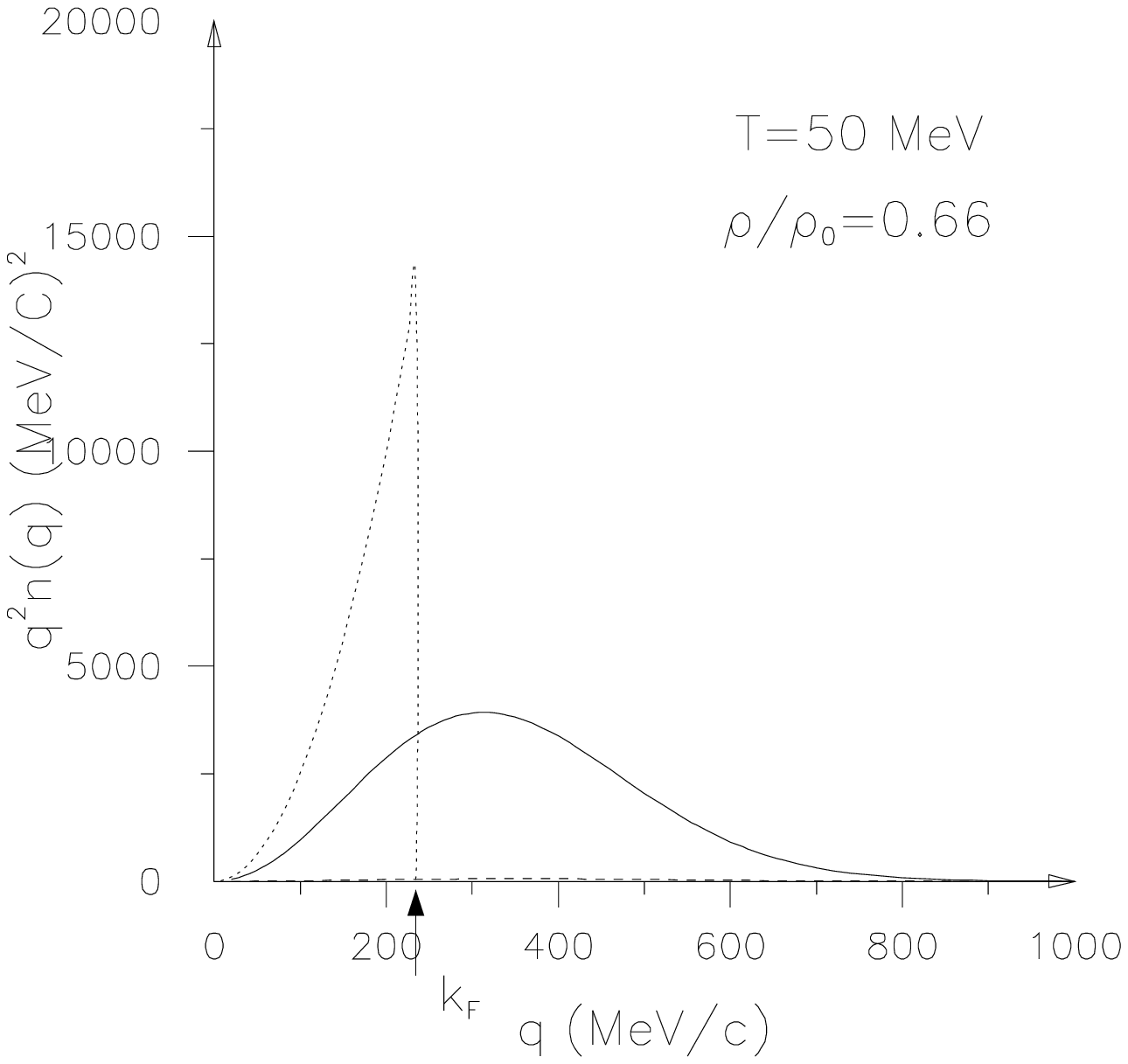,width=4.1cm,height=5.5cm}
&
\epsfig{file=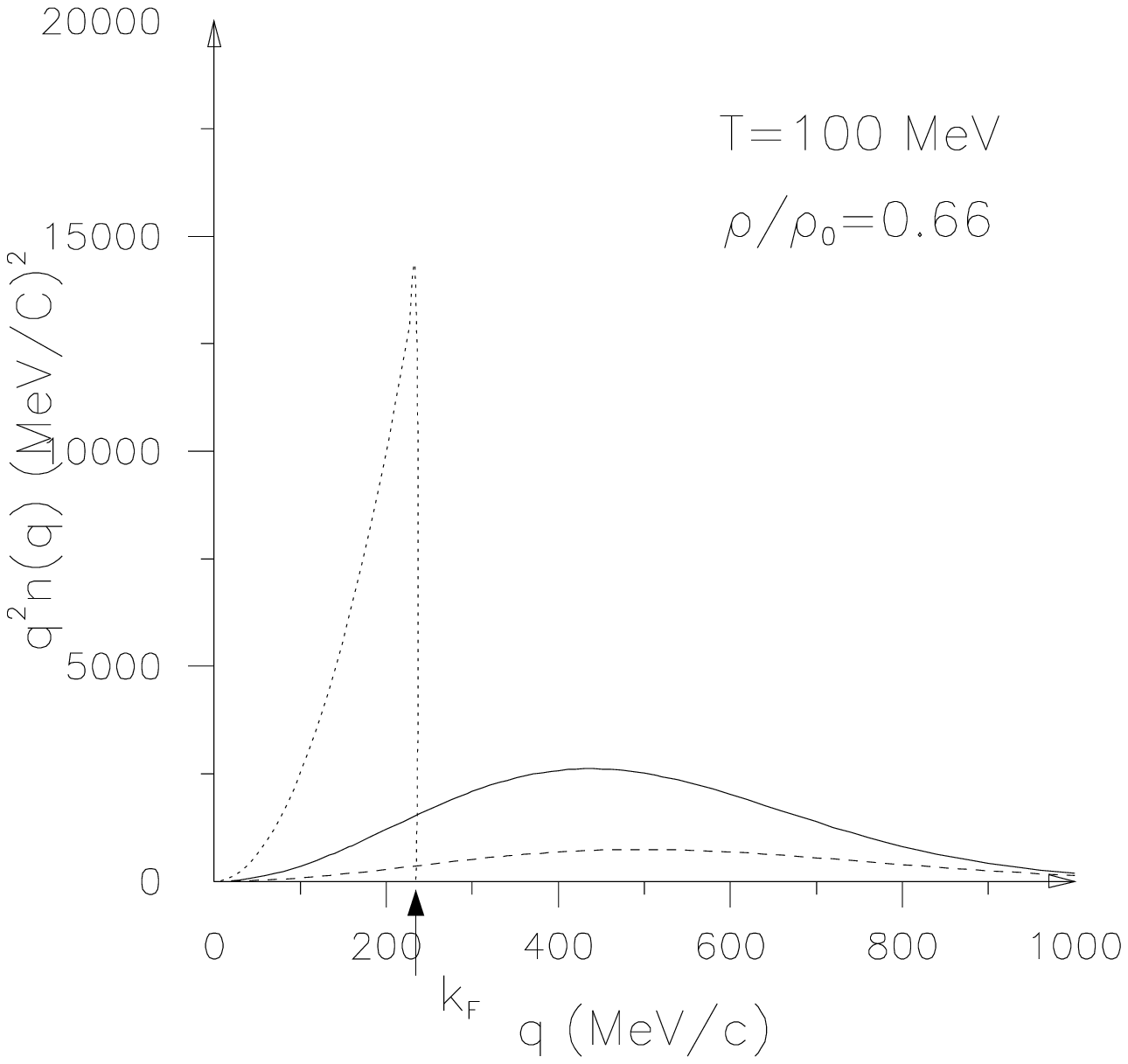,width=4.1cm,height=5.5cm}
&
\epsfig{file=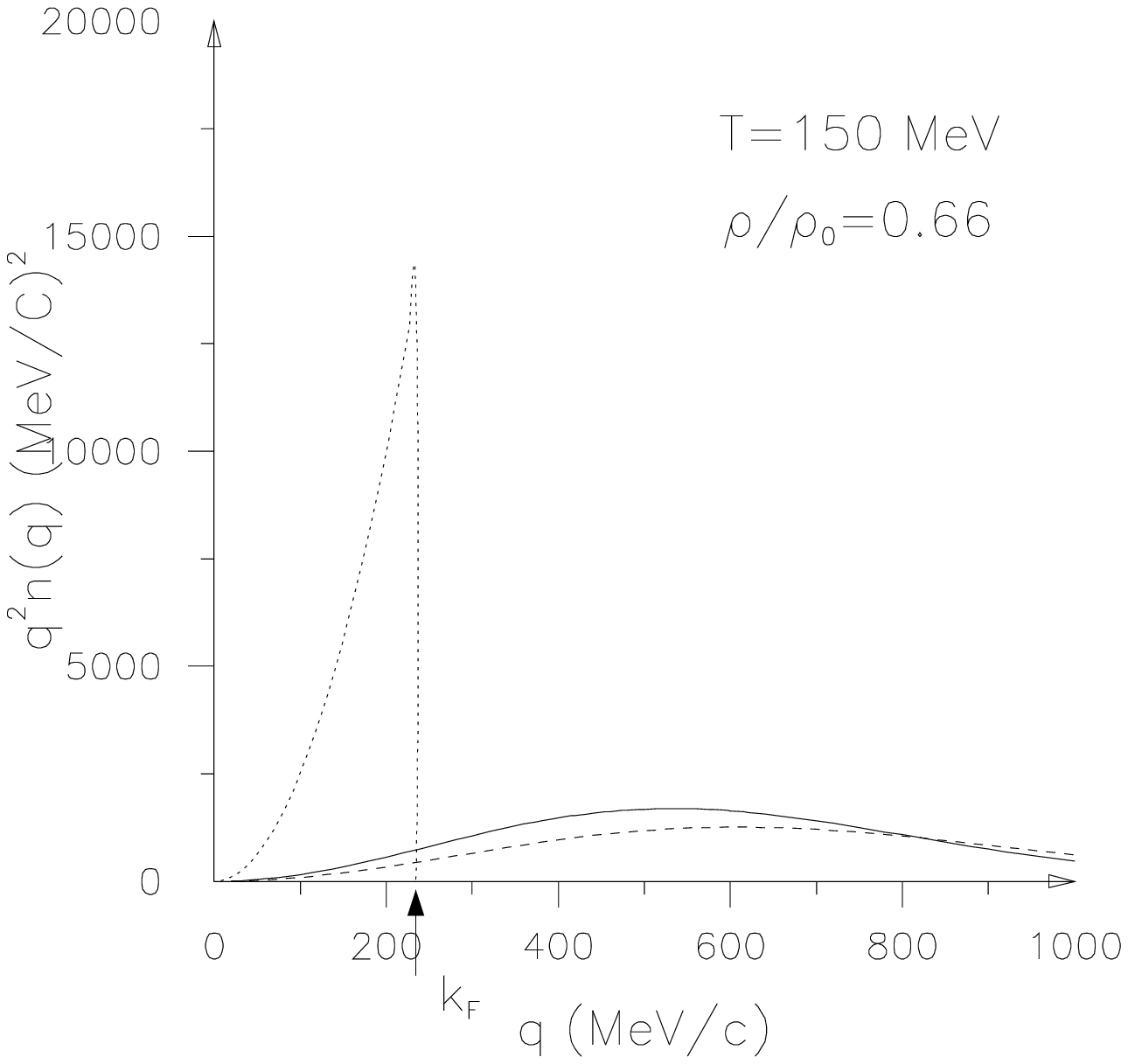,width=4.1cm,height=5.5cm}
\\
\epsfig{file=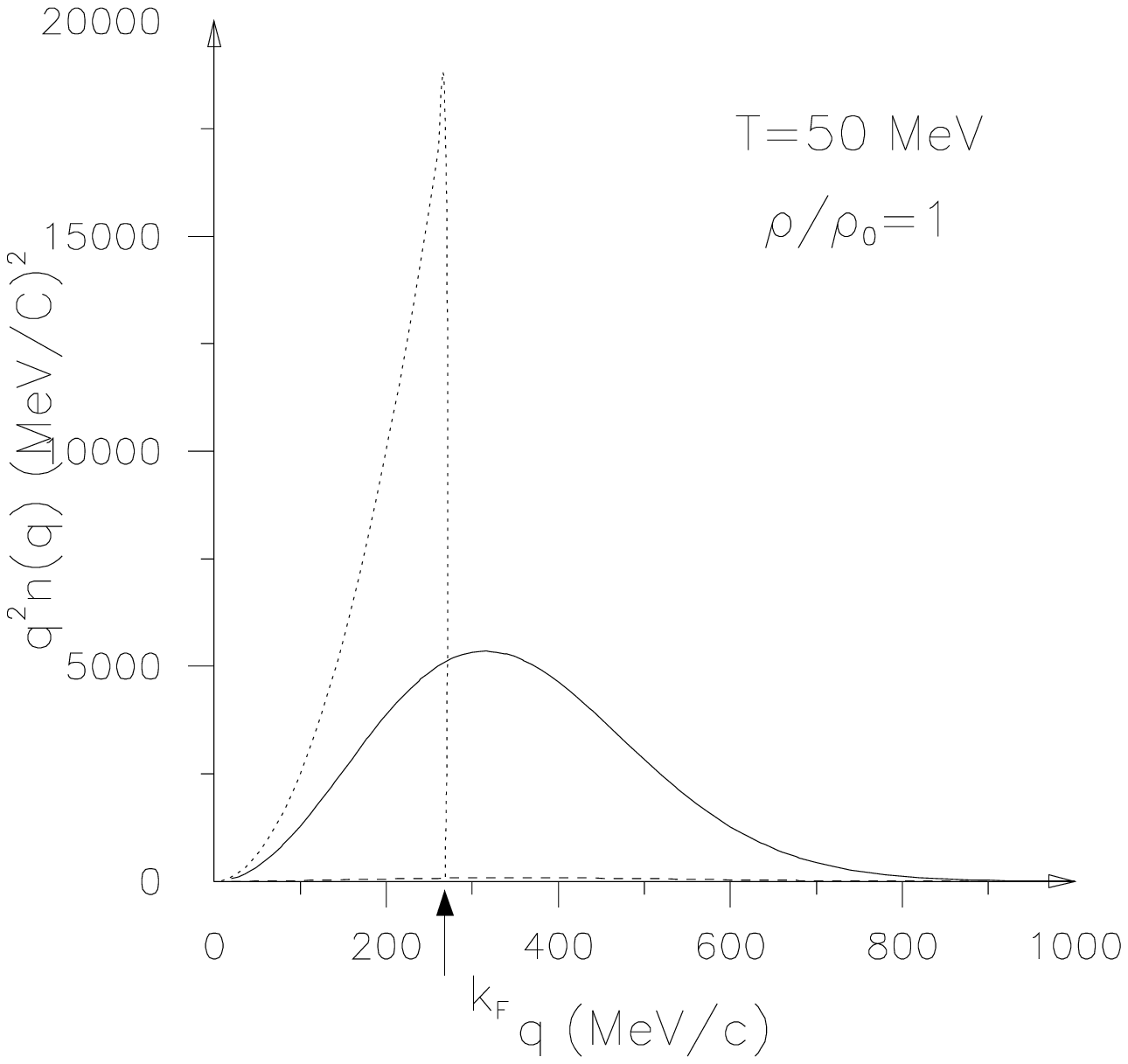,width=4.1cm,height=5.5cm}
&
\epsfig{file=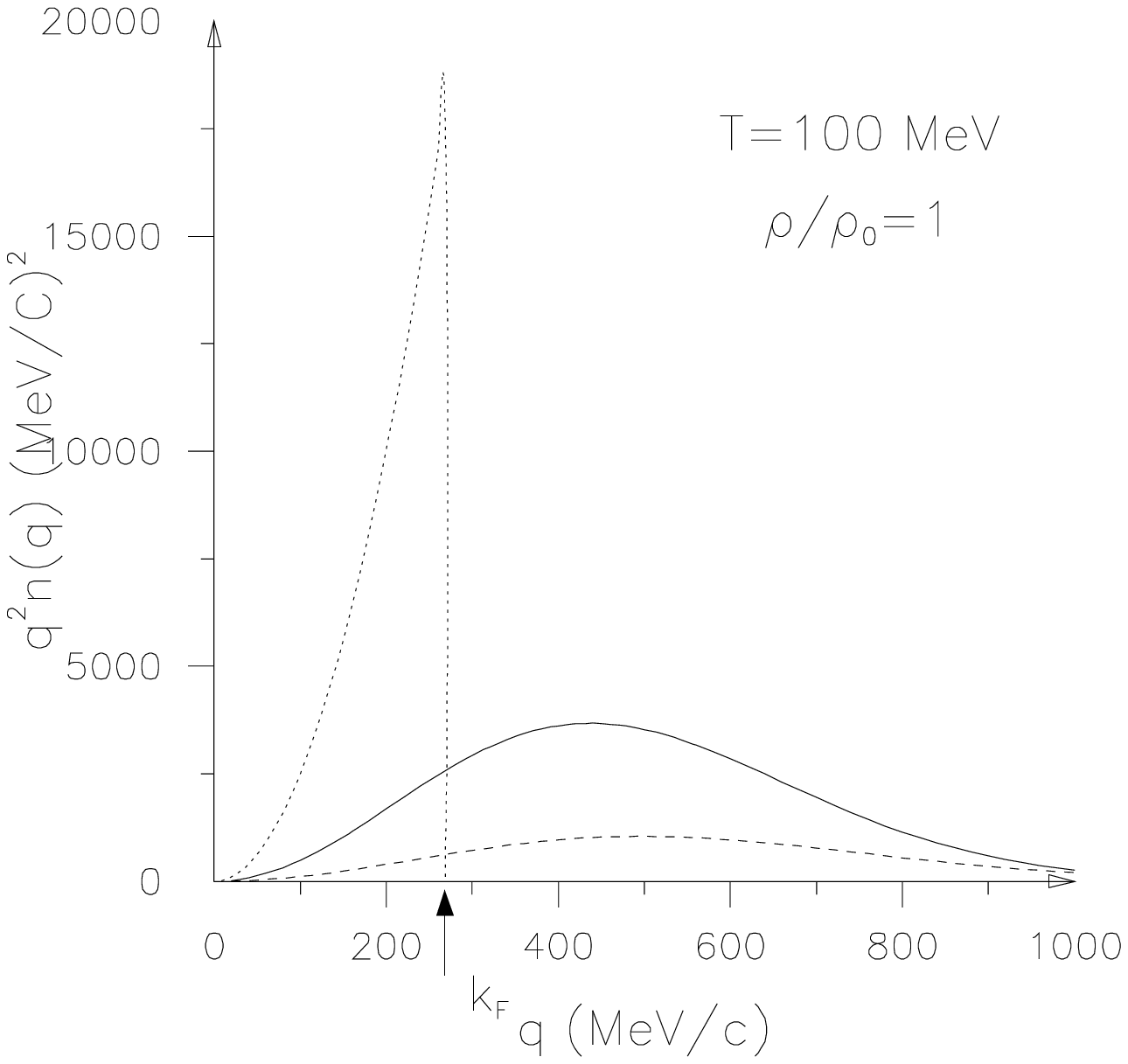,width=4.1cm,height=5.5cm}
&
\epsfig{file=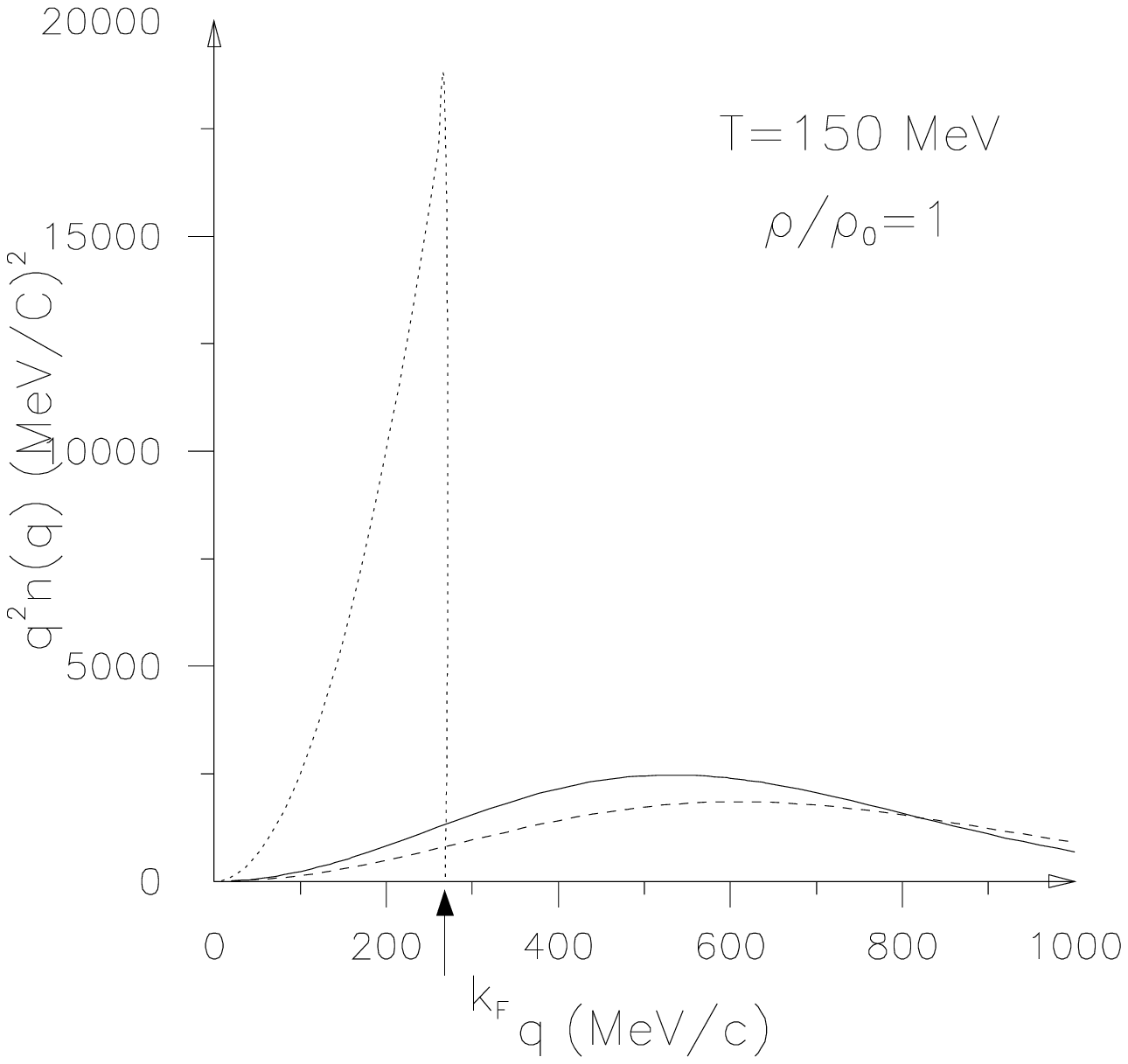,width=4.1cm,height=5.5cm}
  \end{tabular}}
  \caption{Momentum distribution multiplied by $q^2$
of nucleons (solid line) and $\Delta$'s
(dashed line) at different densities and temperatures.
The dotted line corresponds to the Free Fermi Gas at 0 temperature.}
  \label{fig:9}
\end{center}
\end{figure}

Furthermore, at high temperature the dependence upon the density seems 
to become weak and something like an universal behaviour could be suggested.

\section{Discussions}

People speculate about the number of isobars in a hot soup of hadrons
produced in a heavy ion collision. Even assuming that the short time
available in heavy ion collisions is enough to thermalize the baryons
and produce a uniform collective flow, the dynamics discussed in this paper
is not sufficient to enlight the great variety of processes that could
occur at finite temperature. 

The vacuum at finite
temperature is known to contain pions, since these particles are
light. Dey et al \cite{DeElIo-90} have shown that finite temperature will
couple channels with different parity and isotopic spin. For example
the $\rho$ and the $A_1$ meson along with a longitudinal pion mix to
order $T^2$. The poles do not move till the next order $T^4$. In the
same way the nucleon at finite $T$ will couple to the $\Delta_{1/2}$
state and the isobar to the nucleon excited state, $N^*_{3/2}(1520)$
as well as the odd parity isobar $\Delta(1700)$ \cite{ElIo-93}.

Further, if we focus on the relic of a previously realized quark-gluon
condesate, then the strangeness contribution should be relevant,
and investigations on strange hyperons is also suggested
\cite{Br-al-01}.

Coming to more conventional degrees of freedom,
it was shown
by Bedaque \cite{Be-96} in a long letter, using chiral perturbation
theory, that the nucleon mass increases a little, but that the nucleon
acquires a substantial width and the mass is decreased so that the
isobar-nucleon splitting becomes smaller. In chiral limit Bedaque's
result would appear in order $T^4$.

The present considerations seem thus suggest as future perspectives
the extension of the present 1-loop calculation by one side to the study
of other observables in a hadron gas, like for instance effective masses and
widths and also the entropy of the system; and on the other side
to the extension of the present scheme to a richer dynamics encompassing
vector mesons and strange hadrons.

A further interesting development, to be carried out in the future, will
be to extract, from the present formalism, the corresponding behaviour
as an expansion in powers of the temperature.

\section*{Aknowledgements}

J.D. and M.D. acknowledge hospitality at Abdus Salam ICTP and a D.S.T.
research grant no. SP/S2/K18/96, 
Govt. of India for partial support. M.D. acknowledges the kind  
hospitality at  the Dipartimento di Fisica, Universit\`a di Genova, Genova.    


\end{document}